\documentclass[journal]{IEEEtran}
\IEEEoverridecommandlockouts

\usepackage{cite}
\usepackage{amsfonts, amsmath, amsthm, amscd, amssymb, array, bm, bbm, mathtools}
\usepackage{algorithmic}
\usepackage{graphicx}

\def\BibTeX{{\rm B\kern-.05em{\sc i\kern-.025em b}\kern-.08em
    T\kern-.1667em\lower.7ex\hbox{E}\kern-.125emX}}
\usepackage[binary-units=true]{siunitx}
\usepackage{pgfplots}
\usetikzlibrary{arrows,matrix,positioning,shapes.misc,bending,calc}
\pgfplotsset{compat=1.16}
\usepackage[acronym]{glossaries}
\glsdisablehyper
\newacronym{adc}{ADC}{analog-to-digital converter}
\newacronym{awgn}{AWGN}{additive white Gaussian noise}
\newacronym{ct}{CT}{continuous-time}
\newacronym{dt}{DT}{discrete-time}
\newacronym{dtft}{DTFT}{discrete-time \gls{ft}}
\newacronym{ft}{FT}{Fourier transform}
\newacronym{kkt}{KKT}{Karush–Kuhn–Tucker}
\newacronym{lhs}{LHS}{left-hand side}
\newacronym{lmmse}{LMMSE}{linear minimum \gls{mse}}
\newacronym{mse}{MSE}{mean squared error}
\newacronym{mmse}{MMSE}{minimum \gls{mse}}
\newacronym{mmwave}{mmWave}{millimeter wave}
\newacronym{mimo}{MIMO}{multiple-input multiple-output}
\newacronym{pdf}{PDF}{probability density function}
\newacronym{psd}{PSD}{power spectral density}
\newacronym{rhs}{RHS}{right-hand side}
\newacronym{rv}{RV}{random variable}
\newacronym{snr}{SNR}{signal-to-noise ratio}
\newacronym{sar}{SAR}{successive approximation register}
\newacronym{svd}{SVD}{singular value decomposition}
\newacronym{thz}{THz}{terahertz}
\newacronym{wss}{WSS}{wide-sense stationary}
\newacronym{ula}{ULA}{uniform linear array}
\newacronym{pcm}{PCM}{pulse-code modulation}
\usepackage{xcolor}
\usepackage{url}
\usepackage{microtype}
\usepackage{placeins}
\usepackage[hidelinks]{hyperref}
\usepackage{doi}

\DeclareMathAlphabet{\mathsfit}{\encodingdefault}{\sfdefault}{m}{sl}
\SetMathAlphabet{\mathsfit}{bold}{\encodingdefault}{\sfdefault}{bx}{sl}
\DeclareMathOperator*{\argmax}{arg\,max}

\newcommand{\trace}[1]{\mathrm{tr}\left({#1}\right)}
\newcommand{\rank}[1]{\mathrm{rank}\left({#1}\right)}
\newcommand{\mse}[1]{\mathrm{MSE}\left({#1}\right)}
\newcommand{\mat}[1]{\mathbf{#1}}
\newcommand{\matr}[1]{\bm{\mathsf{{#1}}}}

\newcommand{\E}{\mathbb{E}}

\newcommand{\cev}[1]{\reflectbox{\ensuremath{\vec{\reflectbox{\ensuremath{#1}}}}}}

\newtheorem{lemma}{Lemma}
\newtheorem{proposition}{Proposition}
\newtheorem{theorem}{Theorem}
\newtheorem{corollary}{Corollary}

\newcommand{\egc}{e.g., }
\newcommand{\iec}{i.e., }

\newcommand{\wrt}{w.r.t.\ }

\usepackage{atbegshi}
\newcommand\MyOverlayStamp{%
	\begin{tikzpicture}[overlay,remember picture]
					\node (title) at ($(current page.north west)+(0.67in,0)$) [
					anchor=north west,
					align=justify,
					outer sep=0pt,
					font=\fontsize{7}{8}\selectfont,
					text width=7.12in] {This document is a preprint of:
					P. Neuhaus, N. Shlezinger, M. Dörpinghaus, Y. C. Eldar, and G. Fettweis, "Task-Based Analog-to-Digital Converters," in IEEE Trans. Signal Process., vol. 69, pp. 5403-5418, 2021, \doi{10.1109/TSP.2021.3095726}.
					};
					\node (copyright) at ($(current page.south west)+(0.67in,0)$) [
					anchor=south west,
					align=justify,
					outer sep=0pt,
					font=\fontsize{7}{8}\selectfont,
					text width=7.12in] {\textcopyright{} 2021 IEEE. Personal use of this material is permitted. Permission from IEEE must be obtained for all other uses, in any current or future media, including reprinting/republishing this material for advertising or promotional purposes, creating new collective works, for resale or redistribution to servers or lists, or reuse of any copyrighted component of this work in other works.
					};
		\end{tikzpicture}%
}
\AtBeginShipout{\AtBeginShipoutAddToBox{\MyOverlayStamp}}

\begin{document}


\title{Task-Based Analog-to-Digital Converters
\thanks{This work has received funding from the Deutsche Forschungsgemeinschaft (DFG, German Research Foundation) – Project-ID 164481002 – SFB 912, HAEC, the German Federal Ministry of Education and Research (BMBF) (project E4C, contract number 16ME0189), the Benoziyo Endowment Fund for the Advancement of Science, the Estate of Olga Klein -- Astrachan, in part by the European Union’s Horizon 2020 research and innovation program under grant No. 646804-ERC-COG-BNYQ and from the Israel Science Foundation under grant No. 0100101. Computations were performed at the Center for Information Services and High Performance Computing (ZIH) at TU Dresden. This work has been presented in part at the 29th European Signal Processing Conference (EUSIPCO) 2021, Dublin, Ireland \cite{NeuhausEUSIPCO2021a}.}%
\thanks{P. Neuhaus, M. D\"orpinghaus and G. Fettweis are with the Vodafone Chair Mobile Communications Systems, Technische Universit\"at Dresden, 01062 Dresden, Germany (e-mail: \{peter\_friedrich.neuhaus, meik.doerpinghaus, gerhard.fettweis\}@tu-dresden.de).}%
\thanks{N. Shlezinger is with the School of ECE, Ben-Gurion University of the Negev, Be'er-Sheva 84105, Israel (e-mail: nirshl@bgu.ac.il).}%
\thanks{Y. C. Eldar is with the Faculty of Math and CS, Weizmann Institute of Science, Rehovot 7610001, Israel (e-mail:  yonina.eldar@weizmann.ac.il).}%
}

\author{
    \IEEEauthorblockN{
        Peter Neuhaus,~\IEEEmembership{Graduate Student Member,~IEEE,}
        Nir Shlezinger,~\IEEEmembership{Member,~IEEE,}\\
        Meik D\"orpinghaus,~\IEEEmembership{Member,~IEEE,}
        Yonina C. Eldar,~\IEEEmembership{Fellow,~IEEE,}
        and Gerhard Fettweis,~\IEEEmembership{Fellow,~IEEE}
        }
        \vspace{-1.0cm}
}

\maketitle
\pagestyle{plain}
\thispagestyle{plain}

\begin{abstract}
Obtaining digital representations of multivariate continuous-time (CT) signals is a challenge encountered in many signal processing systems. In practice, these signals are often acquired to extract some underlying information, i.e., for a specific task. Employing conventional task-agnostic analog-to-digital converters (ADCs), typically designed to minimize the mean squared error (MSE) in reconstructing the CT input signal, can be costly and energy-inefficient in such cases. In this work, we study task-based ADCs, which are designed to obtain a digital representation of a multivariate CT input process with the goal of recovering an underlying statistically related parameter vector, referred to as \emph{the task}. The proposed system employs analog filtering, uniform sampling, and scalar uniform quantization of the input process before recovering the task vector using a linear digital recovery filter. We optimize the analog and digital filters and derive closed-form expressions for the achievable MSE in recovering the task vector from a set of analog signals when utilizing ADCs with a fixed sampling rate and amplitude resolution. Based on our derivation, we provide guidelines for designing practical acquisition systems subject to a constraint on the bit rate. Our analysis proves that the intuitive approaches of either recovering the task vector solely in digital or designing the analog filter to estimate the task vector are inferior to the proposed joint design. We then consider the recovery of a set of matched filter outputs under a rate budget. We numerically verify our theoretical observations and demonstrate that task-based ADCs substantially outperform analog matched filtering as well as applying the matched filter solely in the digital domain. When acquiring signals for a task under tight bit budgets, we also show that it is often preferable to sample below the Nyquist rate instead of reducing the quantization resolution.
\end{abstract}


\vspace{-0.4cm}
\section{Introduction}\label{sec:intro}
\vspace{-0.1cm}
\Glspl{adc} allow physical signals to be processed using digital hardware. They perform two operations: sampling, i.e., converting a \gls{ct} signal into a discrete set of samples, and quantization, where the continuous-amplitude samples are converted into a finite-bit representation.
Conventional \glspl{adc} are designed to facilitate the recovery of their input signal, where the sampling rate is chosen matched to the bandwidth of the input signal, while the quantizer resolution is set such that the quantization distortion is sufficiently small \cite{eldar2015sampling}. 
In many applications, analog signals are acquired not with the goal of being reconstructed, but for a specific task, e.g., to extract some information from the \gls{ct} input. For instance, \gls{mimo} communications receivers acquire electromagnetic waves measured at their antennas in order to recover a transmitted message or estimate an underlying channel \cite{goldsmith2005wireless}. Radar receivers capture impinging echos in order to identify targets \cite{levanon2004radar}. In such scenarios, utilizing conventional \glspl{adc}, which are designed to recover the analog signals accurately, can be inefficient. Since the power consumption of \glspl{adc} grows with the sampling rate and the quantization resolution \cite{murmann2013energy}, such inefficiency directly leads to increased power consumption. The power consumption of conventional \glspl{adc} is considered a major challenge in beyond 5G systems, which are foreseen to utilize a large number of antennas, i.e., massive \gls{mimo}, as well as large bandwidths in the \gls{mmwave} bands, to meet the ever-increasing demand for higher data rates.

Recently, it was demonstrated that a-priori knowledge about the system task can be exploited to design \emph{task-based quantizers}, which facilitate carrying out the system task while acquiring the input with a limited bit budget \cite{shlezinger2020task}. 
The works \cite{shlezinger2018hardwarelimited,8736805,salamatian2019task,shlezinger2019deep}, which focused on the quantization aspect of analog-to-digital conversion, showed that the distortion induced by low-resolution quantization can be mitigated by accounting for the task. This is achieved by introducing pre-quantization processing, resulting in hybrid analog-digital systems, as commonly utilized in \gls{mimo} systems for the purpose of reducing the number of RF chains \cite{Heath2016,gong2020rf}.
For the sampling operation, such analog processing was shown to facilitate the reconstruction of sub-Nyquist sampled, frequency-sparse analog signals \cite{5419072}, as well as to exploit spatial correlation via joint sampling of multivariate \gls{ct} signals \cite{shlezinger2019joint}. The works \cite{5361485,5419072,shlezinger2019joint} all focused on the reconstruction of the sampled signals, and thus did not consider the presence of a task.

Joint sampling and quantization was investigated in  \cite{8350400}, where the minimum achievable reconstruction distortion under a given rate budget was studied as an indirect source coding problem. These performance bounds can only be achieved by vector quantizers, which are challenging to implement. However, the work \cite{8350400} does not account for the presence of a task, nor does it focus on simple scalar uniform quantizers.
Finally, the recent work \cite{shlezinger2020learning} used deep learning to design task-based \glspl{adc}, including both sampling and quantization, empirically demonstrating the potential gains of such joint designs without providing a theoretical analysis. 

In this work, we consider the design and analysis of an acquisition system for the task of recovering a \emph{linear} function of the observed signals. Such tasks can represent, e.g., channel estimation, as studied in \cite{8736805}, or matched filtering, as considered in our numerical analysis. 
Following \cite{shlezinger2018hardwarelimited,8736805,salamatian2019task, shlezinger2019deep,shlezinger2020task}, we consider a hybrid system with pre-acquisition analog combining and optimize the overall system in light of the task.
However, as opposed to these works, which focused solely on the quantization aspect of acquisition applied to continuous-amplitude vectors, here we consider the complete analog-to-digital conversion procedure, which incorporates both sampling and quantization applied to a set of \gls{ct} continuous-amplitude processes.
In order to allow for a practical implementation of the considered acquisition system,
we focus our analysis on \emph{linear} pre-acquisition mappings, which can be implemented using analog filters, and on \glspl{adc} carrying out uniform sampling and quantization, \iec we consider simple \glspl{adc} and do not allow for complex vector quantizers.

We analytically characterize the minimum achievable \gls{mse} in recovering the desired task from the analog input signals using hybrid acquisition systems with non-overloading quantizers for a given number of \glspl{adc}, sampling rate, and quantization resolution.
We obtain analytical expressions for the analog and digital filters, which achieve this distortion, and identify practical design guidelines for determining the number of \glspl{adc}, the sampling rate, and the quantizer resolution when operating under an overall bit rate constraint.
While optimizing these key \gls{adc} parameters would be desirable when operating under a bit rate constraint, the resulting problem is intractable.
Hence, we numerically search for the \gls{mse}-minimizing \gls{adc} parameterization subject to the bit rate constraint by utilizing the derived closed-form \gls{mse} expressions for a fixed \gls{adc} parameterization.
Our design relies on knowledge of the \emph{statistical} relationship between the observed signals and the desired task.
Hence, it can be applied when one has access to such knowledge during system design.
When such knowledge is only available at the runtime, controllable hardware, as used, \egc in \cite{gong2020rf}, is required.

By jointly studying the complete analog-to-digital conversion procedure in light of the system task, we obtain characterizations and insights which cannot be uncovered using previous works which focused  on task-based quantization.
In particular, our results show that neither recovering the task in the analog domain and subsequently sampling and quantizing the analog \gls{mmse} estimate, nor a fully-digital architecture, which estimates the task solely in the digital domain, minimize the \gls{mse}. Specializing our results to bandlimited signals reveals that it can be preferable to sample below the Nyquist rate to balance the quantization distortion under an overall bit rate budget.
This observation agrees with similar results in \cite{8350400}, which considered joint sampling and lossy source coding.\looseness-10

In our numerical study, we apply the proposed task-based \gls{adc} to the task of estimating the matched filter output of a multi-antenna system, representing a common signal processing task in \gls{mimo} communications and radar. Our numerical results demonstrate the gains of jointly designed task-based \glspl{adc} over the two intuitive approaches of applying the filters solely in the digital or analog domain. In particular, we show that the proposed system can achieve a normalized time-averaged \gls{mse} of $10^{-2}$ using only $\SI{\sim 25}{\percent}$ and $\SI{\sim 8}{\percent}$ of the bit rate budget as compared to analog and digital recovery systems, respectively. Furthermore, we demonstrate that when acquiring bandlimited signals for recovering a task vector under tight bit rate budgets, it is often preferable to sample below the Nyquist rate over reducing the quantization resolution below \SI{2}{bits} per sample.

Compared to our initial results reported in the conference paper \cite{NeuhausEUSIPCO2021a}, we consider arbitrary sampling rates here, and we do not limit the analysis to Nyquist rate sampling of bandlimited signals.
This generalization allows examining a broader range of admissible \gls{adc} parameterizations subject to the considered rate budget compared to \cite{NeuhausEUSIPCO2021a}.
Additional extensions of the current work upon \cite{NeuhausEUSIPCO2021a} include: \hbox{\emph{i}) obtaining} further design guidelines for practical acquisition systems; \hbox{\emph{ii}) investigating} the performance when operating with asymptotically large sampling rates or asymptotically large amplitude resolutions; and \hbox{\emph{iii}) numerically} evaluating the validity of the derived \gls{mse} expressions when employing quantizers with non-zero overload probability and when employing non-dithered quantizers.\looseness-1

The rest of this paper is organized as follows: Section~\ref{sec:system_model} details the model of the proposed task-based acquisition system and formulates the problem. Section~\ref{sec:task_based_adcs} derives the achievable \gls{mse} of task-based \glspl{adc}, designs its components, provides practical design guidelines, and performs an asymptotic performance evaluation.  Section~\ref{sec:num_res} provides a numerical study. Finally, concluding remarks are given in Section~\ref{sec:conclusions}. Detailed proofs of the results presented in the main part of this paper are delegated to the appendix.

Throughout this paper, random quantities are denoted by sans-serif letters, e.g., $\mathsf{a}$, whereas $a$ is a deterministic quantity. Vectors and matrices are denoted by lower and upper case boldface letters, e.g., $\mat{a}$ and $\mat{A}$, respectively. $\mat{I}_K$ is the $K\times K$ identity matrix. We use $j$, $\ast$, $\mathrm{tr}(\cdot)$, $(\cdot)^{\dagger}$, $\mathcal{F} \left\lbrace \cdot \right\rbrace$, and $\mathbb{E}\{\cdot\}$ to denote the imaginary unit, convolution, the trace operator, pseudo-inverse, \gls{ft}, and stochastic expectation, respectively. Furthermore, we use the shorthand notation $\cev{x}(t) = x(-t)$ for time reversal. The sets of natural, integer, real and complex numbers are written as $\mathbb{N}$, $\mathbb{Z}$, $\mathbb{R}$ and $\mathbb{C}$, respectively.\looseness-10

\vspace{-0.2cm}
\section{System Model}\label{sec:system_model}
\vspace{-0.1cm}
\begin{figure*}
	\centering
	\includegraphics[width=0.81\textwidth]{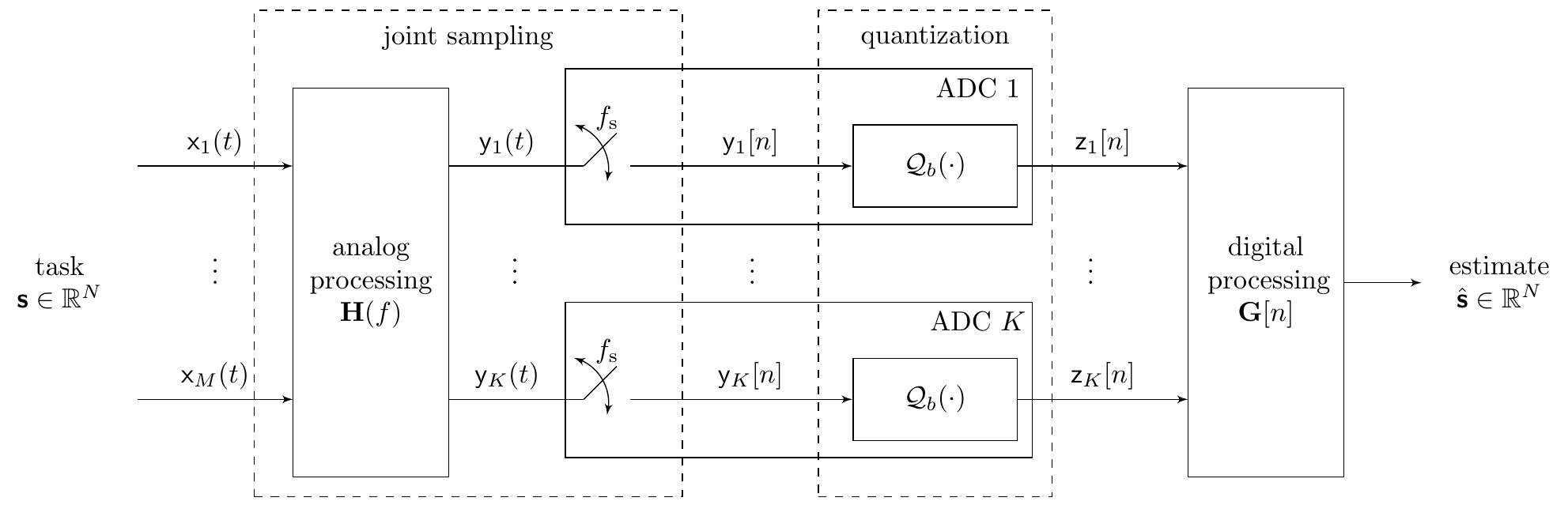}
	\caption{Overview of the system model. The task is to estimate the random vector $\matr{s} \in \mathbb{R}^N$. The digital task estimate is denoted as $\hat{\matr{s}} \in \mathbb{R}^N$.\vspace{-0.2cm}}
	\label{fig:system_model}
\end{figure*}
In this work, we consider a hybrid analog-digital acquisition system model, as illustrated in Fig.~\ref{fig:system_model}. The system is comprised of a joint sampling block, consisting of a multivariate analog filter and a set of uniform samplers and a set of uniform quantizers, and a digital processor. This division aims to facilitate system design. In a practical system, the analog filter requires dedicated  circuitry, while the uniform samplers and quantizers correspond to conventional scalar \glspl{adc}. The overall system is optimized for recovering a task vector from a set of analog signals.

In the following, we formulate the problem of jointly designing the components of the hybrid acquisition system. We first review the signal model relating the observed signals and the task vector in Subsection~\ref{sec:Signal}. Then, we model the joint sampling and quantization operations  in Subsections \ref{sec:joint_sampling} and \ref{sec:quantization}, respectively. Based on these models, we formulate the problem of optimizing the system in Subsection \ref{sec:problem_formulation}.

\vspace{-0.2cm}
\subsection{Signal Model}
\label{sec:Signal} 
\vspace{-0.1cm}
The observations comprise a set of $M \geq 1$ \gls{ct} signals $\left\lbrace \mathsf{x}_m(t) \right\rbrace_{m=1}^{M}, t \in \mathbb{R}$, as illustrated in Fig.~\ref{fig:system_model}. We model these signals as zero-mean jointly \gls{wss} random processes with finite variances, and let  $\mat{C}_{\matr{x}}(f) \in \mathbb{C}^{M\times M}, f \in \mathbb{R},$ denote their joint \gls{psd}. This set of analog signals is acquired for the task of estimating a zero-mean random vector $\matr{s} \in \mathbb{R}^N$, referred to as {\em the task}.
We assume a known statistical relationship between the \gls{ct} random input signals $\left\lbrace \mathsf{x}_m(t) \right\rbrace$ and the task vector $\matr{s}$.

In particular, we assume that the \gls{mmse} estimate of $\matr{s}$ from $\left\lbrace \mathsf{x}_m(t) \right\rbrace_{m=1}^{M}$ takes a linear form. By defining $\matr{x}(t) \triangleq [\mathsf{x}_1(t), \ldots, \mathsf{x}_M(t)]^T$, this assumption implies that there exists a multivariate filter $\mat{\Gamma}(t) \in \mathbb{R}^{N\times M}$ such that
\vspace{-0.1cm}
\begin{equation}
    \tilde{\matr{s}} \triangleq \E\left\{\matr{s} \Big\vert \left\lbrace \mathsf{x}_m(t) \right\rbrace_{m=1}^{M} \right\} = \int_{\mathbb{R}} \cev{\mat{\Gamma}}(t) \matr{x}(t) \mathrm{d}t = \left( \mat{\Gamma} * \matr{x} \right)(0).
    \vspace{-0.1cm}
    \label{eq:def_Gamma_t}
\end{equation}
The resulting formulation allows modeling tasks which can be expressed as linear functions of the observed signals. Such problems of recovering linear functions of analog signals arise, e.g., in matched-filtering based processing, commonly used in \gls{mimo} radar \cite[Ch.~2]{levanon2004radar} and communications \cite{biglieri2007mimo}, as well as channel estimation in rich scattering environments \cite[Sec.~3.1]{marzetta2016fundamentals}.
Moreover, note that defining the task with respect to (w.r.t.) to the time instance $t=0$ is not a limitation, because we can always reformulate \eqref{eq:def_Gamma_t} \wrt $\tilde{\mat{\Gamma}}(t) \triangleq \mat{\Gamma}(t-t_0)$, for some $t_0 \in \mathbb{R}$.\looseness-1

\vspace{-0.2cm}
\subsection{Joint Sampling Operation}
\label{sec:joint_sampling}
\vspace{-0.1cm} 
The acquired analog signals are first converted into a set of discrete-time sequences via \emph{joint sampling}. Joint sampling is a framework for sampling a set of \gls{ct} signals, allowing to exploit their spatial correlation using multivariate analog filters \cite{shlezinger2019joint}. 
In particular, the $M$ \gls{ct} input signals are filtered by a multivariate analog filter $\mat{H}( f ) \in \mathbb{R}^{K \times M}, f \in \mathbb{R}$, resulting in $K \leq M$ \gls{ct} signals $\{\mathsf{y}_k(t) \}_{k=1}^K$, obtained by 
\vspace{-0.1cm}
\begin{equation}
	\mathsf{y}_k(t) = \sum_{m=1}^{M} \left(h_{k,m}* \mathsf{x}_m\right)(t),  \quad \forall k \in \mathcal{K} \triangleq \{1,\ldots,K\}.
	\label{eqn:FiltOutput}
\vspace{-0.1cm}
\end{equation}
 In \eqref{eqn:FiltOutput}, $h_{k,m}(t)$ denotes the impulse response of a scalar filter which is the inverse \gls{ft} of $\left[ \mat{H} \right]_{k,m}(f)$, \iec $h_{k,m}(t) \triangleq \mathcal{F}^{-1}\{\left[ \mat{H} \right]_{k,m}(f)\}.$

The outputs $\left\lbrace \mathsf{y}_k(t) \right\rbrace_{k=1}^K$ of the analog filter are sampled by $K$ identical uniform samplers, each with sampling rate $f_\mathrm{s}$, i.e., samples are spaced by $T_\mathrm{s} = \frac{1}{f_\mathrm{s}}$. This yields the set of $K$ discrete-time sequences $\left\lbrace \mathsf{y}_k[n] \right\rbrace_{k=1}^K$ where $\mathsf{y}_k[n] \triangleq T_\mathrm{s} \mathsf{y}_k(n T_\mathrm{s})$.

\vspace{-0.2cm}
\subsection{Quantization Operation}
\label{sec:quantization}
\vspace{-0.1cm}
The sampled signals $\left\lbrace \mathsf{y}_k[n] \right\rbrace_{k=1}^K $ are subsequently quantized by $K$ identical uniform scalar mid-rise quantizers with an amplitude resolution of $b$ bits. Such quantizers produce up to $2^b$ distinct output values. The (one-sided) support of the quantizers, also referred to as the {\em dynamic range}, is denoted as $\gamma > 0$. The quantization step size is hence given by $\Delta = \frac{2 \gamma}{2^b}$, and the mid-rise quantization function is then defined as
\begin{equation}
	q_b\left( x' \right) \triangleq \begin{cases}
		\Delta \left( \left\lfloor \frac{x'}{\Delta} \right\rfloor + \frac{1}{2} \right), & \mathrm{for~} \vert x' \vert < \gamma\\
		\mathrm{sign} \left( x' \right) \left( \gamma - \frac{\Delta}{2} \right), & \mathrm{otherwise,}
	\end{cases}
	\label{eq:def_mid-rise_quantization}
\end{equation}
where $\lfloor \cdot \rfloor$ denotes rounding to the next smaller integer.

To obtain an analytically tractable model for such non-linear quantizers, which accurately represents their operation in a broad range of setups, we focus on nonsubtractive dithered quantizers operating within their dynamic range as done in \cite{shlezinger2018hardwarelimited}. The resulting quantization model is detailed in the following:

\subsubsection{Nonsubtractive Dithered Quantizers}
In our analysis of task-based \glspl{adc}, we model the non-linear quantizers as implementing \emph{nonsubtractive dithered quantization}. In such continuous-to-discrete mappings, an additional signal referred to as \emph{dither} is added to the quantizer input prior to quantization \cite{gray1993ditheredQuantizers}. Unlike subtractive dithered quantizers \cite{zamir1992universal}, here, the dither is not subtracted, i.e., compensated, after quantization. The quantizer outputs are given by
\begin{equation}
    \mathsf{z}_k[n] = \mathcal{Q}_{b}\left( \mathsf{y}_k[n] \right) = q_b\left( \mathsf{y}_k[n] + \mathsf{w}_k[n] \right), \qquad k \in \mathcal{K},
\end{equation}
where $\left\lbrace \mathsf{w}_k[n] \right\rbrace_{k=1}^K$ denotes the dither random process, which is independent and identically distributed (i.i.d.) in time and space, and mutually independent of the input process. In order to obtain a tractable equivalent quantizer model detailed later in this section, the \gls{pdf} of the dither random process is chosen as a triangular function with a width of $2 \Delta$, \iec it is given by
\begin{equation}
    p_{\mathsf{w}} (w) = \begin{cases}
        \frac{1}{\Delta}\left(1-\frac{|w|}{\Delta}\right), & \mathrm{for~} |w| \leq \Delta\\
        0, & \mathrm{otherwise.}
    \end{cases}
    \label{eq:def_pdf_dither}
\end{equation}

For non-overloading quantizers, i.e., for inputs that never exceed the quantizer's dynamic range, the specific choice of a triangular \gls{pdf} ensures that the first and second moments of the quantization error are independent of the input while minimizing the second \cite[Sec. III.C]{wannamaker2000nonsubtractiveDither}.
While this model allows rigorously characterizing the \gls{mse}-minimizing task-based acquisition system in Subsection \ref{sec:main_results}, the resulting system is also applicable when using conventional non-dithered quantizers as discussed in Subsection \ref{subsec:discussion} and numerically verified in Subsection \ref{sec:num_res_eval_framework}.\looseness-1

\subsubsection{Dynamic Range}
Quantizers are typically required to operate within their dynamic range to yield distinguishable digital representations of different inputs \cite{gray1998quantization}. Consequently, the overload probability, i.e., the probability that the input's magnitude exceeds the dynamic range, has to be negligible. Hence, the dynamic range $\gamma$ is chosen to a multiple $\eta$ of the largest standard deviation of the quantizer inputs, i.e.,
\begin{equation}
    \gamma^2 =  \eta^2 \max_{k \in \mathcal{K}} \mathbb{E}\{ \tilde{\mathsf{y}}_k^2[n] \},
    \label{eq:def_dynamic_range_squared}
\end{equation}
with $\tilde{\mathsf{y}}_k[n] \triangleq \mathsf{y}_k[n] + \mathsf{w}_k[n]$.
Using Chebychev's inequality, the overload probability is now upper-bounded by \cite[eq.~(5-88)]{papoulis2001probability}
\begin{equation}
    \mathrm{Pr} \left\lbrace \vert \tilde{\mathsf{y}}_k[n] \vert \geq \gamma  \right\rbrace \leq \frac{ \mathbb{E}\{ \tilde{\mathsf{y}}_k^2[n] \} }{ \eta^2 \max_{k' \in \mathcal{K}} \mathbb{E}\{ \tilde{\mathsf{y}}_{k'}^2[n] \} } \leq \frac{1}{\eta^2}. 
    \label{eqn:Overload}
\end{equation}

The overload probability bound in \eqref{eqn:Overload} holds regardless of the distribution of $\tilde{\mathsf{y}}_k[n]$, and tighter bounds can be obtained by accounting for its distribution. For instance, when $\tilde{\mathsf{y}}_k[n]$ is Gaussian, setting $\eta = 2$ yields an overload probability of less than \SI{5}{\percent}, compared to the  bound of \SI{25}{\percent} obtained using \eqref{eqn:Overload}.

\subsubsection{Equivalent Quantizer Model}\label{sec:equivalent_quantizer_model}
For the considered dithered quantizer model with zero overload probability, i.e., $\mathrm{Pr} \left\lbrace \vert \tilde{\mathsf{y}}_k[n] \vert \geq \gamma  \right\rbrace = 0$, 
it follows from \cite[Th.~2]{gray1993ditheredQuantizers} that the quantizer output can be written as
\begin{equation}
    \matr{z}[n] = \matr{y}[n] + \matr{e}[n],
    \label{eq:def_z_n}
\end{equation}
where $\matr{z}[n] \triangleq [\mathsf{z}_1[n], \ldots, \mathsf{z}_K[n]]^T$ and $\matr{y}[n] \triangleq [\mathsf{y}_1[n], \ldots, \mathsf{y}_K[n]]^T$.
The quantization error $\matr{e}[n] \in \mathbb{R}^{K}$ in \eqref{eq:def_z_n} is:   \emph{i})  uncorrelated with  $\matr{y}[n]$; \emph{ii}) comprised of uncorrelated entries, and its autocorrelation function  is  
\begin{equation}
    \mat{C}_{\matr{e}}[l] = \mathbb{E} \left\lbrace \matr{e}[n+l] \matr{e}^T[n] \right\rbrace = \frac{\Delta^2}{4} \mat{I}_K \delta[l],
    \label{eq:def_C_e}
\end{equation}
where $\delta[n]$ denotes the Kronecker delta function. The resulting equivalent quantizer model from \eqref{eq:def_z_n} and \eqref{eq:def_C_e} also holds approximately when using conventional non-dithered quantizers as discussed in Subsection \ref{subsec:discussion} and numerically verified in Subsection \ref{sec:num_res_eval_framework}.

\vspace{-0.2cm}
\subsection{Problem Formulation } \label{sec:problem_formulation}
\vspace{-0.1cm}
Our goal is to design the task-based acquisition system illustrated in Fig.~\ref{fig:system_model} to recover the task vector $\matr{s}$ accurately. We aim to optimize the system's components jointly, i.e., its analog filtering and digital processing, to minimize the \gls{mse} between the task vector $\matr{s}$ and its estimate $\hat{\matr{s}}$, under a constraint on the maximum bit rate $ K \cdot f_\mathrm{s} \cdot b \leq R$. Note that the bit rate budget $R$ represents the maximal amount of bits the system can generate per second, thus relating to the implementation complexity and the power consumption.

Here, we focus on digital recovery using linear filters; namely, the digital processing consists of a multivariate filter $\mat{G}[n]  \in \mathbb{R}^{N \times K}$, such that the task estimate is given by
\begin{equation}
    \hat{\matr{s}} = \sum_{n \in \mathbb{Z}} \cev{\mat{G}}[n] \matr{z}[n] = \left( \mat{G} \ast \matr{z} \right)[0].
    \label{eq:def_s_hat}
    \vspace{-0.1cm}
\end{equation}
Hence, our goal is to solve the following minimization problem
\begin{equation}
	\mathop{\min}\limits_{\mat{H}(f), \mat{G}[n]} \mathbb{E} \big\lbrace \big\Vert \matr{s} - \hat{\matr{s}} \big\Vert^2 \big\rbrace, \quad \mathrm{s.t.} \quad  K \cdot f_\mathrm{s} \cdot b \leq R,
	\label{eqn:Objective}
\end{equation}
where the expectation is computed \wrt the joint distribution of $\matr{s}$ and $\hat{\matr{s}}$.

We allow the input dimensionality $M$ to take any positive integer value, \iec the input can be a scalar signal. In addition, we focus on systems where the number of \glspl{adc} is not larger than the number of input signals, i.e., $K \leq M$. While this setup represents a broad range of  architectures of practical interest, it does not accommodate acquisition systems based on signal expansion, as used in some generalized sampling mechanisms \cite{eldar2015sampling}, as well as feedback-based schemes, such as Sigma-Delta \glspl{adc} \cite{aziz1996overview}.

The \gls{mse} $\mathbb{E} \lbrace \Vert \matr{s} - \hat{\matr{s}} \Vert^2 \rbrace$ in estimating the task vector $\matr{s}$ can always be decomposed as \cite[eq.~(49)]{wolf1970transmission}
\begin{equation}
    \mathbb{E} \left\lbrace \left\Vert \matr{s} - \hat{\matr{s}} \right\Vert^2 \right\rbrace
    = \mathbb{E} \left\lbrace \left\Vert \matr{s} - \tilde{\matr{s}} \right\Vert^2 \right\rbrace + \mathbb{E} \left\lbrace \left\Vert \tilde{\matr{s}} - \hat{\matr{s}} \right\Vert^2 \right\rbrace,
    \label{eq:mse_decomposition_wolf}
\end{equation}
\iec as the sum of the analog \gls{mmse} and the distortion \wrt the analog \gls{mmse} estimate $\tilde{\matr{s}}$.
Note that the latter depends on the proposed task-based \gls{adc}, whereas the former does not.
Consequently, the \gls{mse} in \eqref{eqn:Objective} can be replaced with $\mathbb{E} \left\lbrace \left\Vert \tilde{\matr{s}} - \hat{\matr{s}} \right\Vert^2 \right\rbrace$, i.e., the error \wrt the analog \gls{mmse} estimate $\tilde{\matr{s}}$ (cf.~\eqref{eq:def_Gamma_t}), as we do in our analysis in the following section.

\vspace{-0.2cm}
\section{Task-Based ADCs}
\label{sec:task_based_adcs}
\vspace{-0.1cm}
In this section, we jointly optimize the components of the task-based acquisition system detailed in the previous section in light of the problem stated in Subsection \ref{sec:problem_formulation}. Directly solving \eqref{eqn:Objective} is challenging due to the dependency of the system's components on the quantities $K$, $f_{\rm s}$, and $b$.
Thus, we tackle this optimization problem by first minimizing the objective of \eqref{eqn:Objective}  for a {\em fixed \gls{adc} configuration}, i.e., assuming that the triplet $(K, f_\mathrm{s}, b)$ is given. Under this assumption, we jointly optimize the analog filter $\mat{H}(f)$, and the digital filter $\mat{G}[n]$ in Subsection \ref{sec:main_results}. While the resulting derivation does not result in a tractable characterization of the optimal triplet $(K^{\mathrm{o}}, f_\mathrm{s}^{\mathrm{o}}, b^{\mathrm{o}})$, it sheds light on the structure and the behavior of the \gls{mse}-minimizing task-based acquisition system, as we show in Subsections \ref{sec:practical_design_guidelines}-\ref{sec:asymptotic_analysis}. In addition, our analytical derivation facilitates numerical optimization of the \gls{adc} parameter triplets, as we discuss in Subsection \ref{subsec:discussion} and demonstrate in Subsection~\ref{sec:num_res_rate_budget}.\looseness-1

\vspace{-0.2cm}
\subsection{Main Results}\label{sec:main_results}
\vspace{-0.1cm}
Here, we characterize the task-based acquisition system which minimizes \eqref{eqn:Objective} under a fixed \gls{adc} configuration. To this aim, we fix the number of \glspl{adc} $K$, the sampling rate $f_{\rm s}$, and the number of quantizer bits $b$.
Under this setting, we first identify the \gls{mse}-minimizing  digital recovery filter $\mat{G}^\mathrm{o}[n]$ for a given analog filter $\mat{H}(f)$, as stated in the following proposition:
\begin{proposition} \label{th:optimal_digital_filter}
    For the considered system model as described in Sec.~\ref{sec:system_model}, a given analog filter $\mat{H}(f)$, a fixed number of \glspl{adc} $K$, a fixed sampling rate $f_\mathrm{s}$, and a given quantizer resolution $b$, the \gls{mse}-minimizing linear digital recovery filter is given by
    \begin{align}
        \label{eq:theorem_optimal_G_omega}
        \mat{G}^{\mathrm{o}}[n] & = T_\mathrm{s} \int_{-\frac{f_\mathrm{s}}{2}}^{\frac{f_\mathrm{s}}{2}} \sum_{k \in \mathbb{Z}} \mat{\Gamma}\left(f - k f_\mathrm{s}\right) \mat{C}_{\matr{x}} \left(f - k f_\mathrm{s}\right) \nonumber \\
        & \quad \times \mat{H}^H\left(f - k f_\mathrm{s}\right) \mat{C}_{\matr{z}}^{-1}(e^{j 2 \pi f T_\mathrm{s}}) \, e^{j 2 \pi f n T_\mathrm{s}} \mathrm{d} f,
    \end{align}
    where $\left[ \mat{\Gamma} \right]_{n,m}(f)$ is  the \gls{ft} of $\left[ \mat{\Gamma} \right]_{n,m}(t)$, and
    \begin{IEEEeqnarray}{lCl}
        \mat{C}_{\matr{z}}(e^{j 2 \pi f T_\mathrm{s}}) & = & T_\mathrm{s} \sum_{k \in \mathbb{Z}} \mat{H}\left( f - k f_\mathrm{s} \right) \mat{C}_{\matr{x}} \left( f - k f_\mathrm{s} \right) \nonumber \\
        && \quad \times \mat{H}^H\left( f - k f_\mathrm{s} \right) + \frac{\Delta^2}{4}\mat{I}_K.
    \end{IEEEeqnarray}
    Furthermore, the resulting minimum achievable \gls{mse} is given by\looseness-1
    \begin{IEEEeqnarray}{ll}
        & \mse{\mat{H}(f)} = \E \left\lbrace \Vert \tilde{\matr{s}} \Vert^2  \right\rbrace \nonumber \\
        & \quad - \mathrm{tr} \left( T_\mathrm{s} \int_{-\frac{f_\mathrm{s}}{2}}^{\frac{f_\mathrm{s}}{2}} \mat{S}(f)  \mat{C}_{\matr{z}}^{-1}(e^{j 2 \pi f T_\mathrm{s}}) \mat{S}^H(f) \, \mathrm{d} f \right),
        \label{eq:theorem_mse_h_omega_general}
    \end{IEEEeqnarray}
    where 
    \begin{equation}
        \label{eq:def_S_of_f}
        \mat{S}(f) \triangleq \sum_{k \in \mathbb{Z}} \mat{\Gamma}\left(f - k f_\mathrm{s}\right) \mat{C}_{\matr{x}} \left(f - k f_\mathrm{s}\right)  \mat{H}^H\left(f - k f_\mathrm{s}\right).
    \end{equation}
\end{proposition}
\begin{IEEEproof}
    The proof is provided in Appendix \ref{sec:appendix_A}.
\end{IEEEproof}

\smallskip
After obtaining the minimum achievable \gls{mse} for a given analog filter $\mat{H}(f)$ for fixed $K$, $f_\mathrm{s}$, and $b$, we proceed to find the analog filter $\mat{H}(f)$, which minimizes the \gls{mse} in \eqref{eq:theorem_mse_h_omega_general}. 
In particular, we derive the analog filter for bandlimited inputs, where there exists a $\Upsilon\in \mathbb{N}$ such that $\left[\mat{C}_\mat{x}(f)\right]_{i,j} = 0$ for all $|f| > (\Upsilon+\frac{1}{2}) f_{\rm s}$, for all $i,j \in \mathcal{M} \triangleq \{1,\ldots,M\}$.
While our following derivation rigorously holds for bandlimited signals, note that we do not assume Nyquist rate sampling. Furthermore, as the inputs $\{\mathsf{x}_m(t)\}_{m=1}^M$ have finite variance, it follows that the entries of the \gls{psd} matrix must decay as $|f|$ grows, and thus purely non-bandlimited signals can be represented with arbitrary accuracy by considering bandlimited signals with a sufficiently large $\Upsilon$.
Consequently, assuming bandlimited inputs allows to be mathematically rigorous while not limiting the practical application of the results, as the band limitation can be arbitrarily large.

By defining  $\bar{M} \triangleq (2 \Upsilon + 1)M$ and 
\begin{align}
    \label{eq:def_Gamma_bar}
     \bar{\mat{\Gamma}}(f) & \triangleq \left[ \mat{\Gamma}_{-\Upsilon}(f) \mat{L}_{-\Upsilon}(f), \ldots, \mat{\Gamma}_{+\Upsilon}(f)  \mat{L}_{+\Upsilon}(f) \right], \\
     \label{eq:def_H_bar}
     \bar{\mat{H}}(f) & \triangleq \left[ \mat{H}_{-\Upsilon}(f)  \mat{L}_{-\Upsilon}(f), \ldots, \mat{H}_{+\Upsilon}(f)  \mat{L}_{+\Upsilon}(f) \right],
\end{align}
where  $\mat{H}_k(f)  \triangleq  \mat{H}(f - k f_\mathrm{s})$, $\mat{L}_k(f) \triangleq  \mat{C}_{\matr{x}}^{1/2}(f - k f_\mathrm{s})$, and $\mat{\Gamma}_k(f)  \triangleq  \mat{\Gamma}(f - k f_\mathrm{s})$, we can characterize the \gls{mse}-minimizing analog filter as stated in the following theorem:
\begin{theorem} \label{th:optimal_pre_filter}
    For the considered system model as described in Sec.~\ref{sec:system_model}, a fixed number of \glspl{adc} $K$, a fixed sampling rate $f_\mathrm{s}$, and a given quantizer resolution $b$, the \gls{mse}-minimizing analog filter $\mat{H}^{\mathrm{o}}(f) \in \mathbb{C}^{K \times M}$ is characterized for each $f \in \big[ -\frac{f_\mathrm{s}}{2}, \frac{f_\mathrm{s}}{2} \big]$ by\looseness-1
    \begin{equation} 
        \label{eq:theorem_H_opt}
        \bar{\mat{H}}^{\mathrm{o}}(f)  =\! \bar{\mat{U}}_{\mat{H}}(f) \bar{\mat{\Sigma}}_{\mat{H}}(f) \bar{\mat{V}}^H_{\mat{H}}(f).
    \end{equation} 
    Specifically, \eqref{eq:theorem_H_opt} is comprised of:
    \begin{itemize}
        \item $\bar{\mat{V}}_{\mat{H}}(f) \in \mathbb{C}^{\bar{M} \times \bar{M}}$ is the matrix of right-singular vectors of $\bar{\mat{\Gamma}}(f)$, which is defined in \eqref{eq:def_Gamma_bar}. 
        \item $\bar{\mat{\Sigma}}_{\mat{H}}(f) \in \mathbb{R}^{K \times \bar{M}}$ is  diagonal with diagonal entries
        \begin{equation}
            \label{eq:theorem_1_def_Sigma}
            \left[\bar{\mat{\Sigma}}_{\mat{H}}(f)\right]_{i,i} = \frac{1}{2^{b}} \sqrt{ \left( \zeta \sigma_{\bar{\mat{\Gamma}},i}(f) - 1 \right)^+},    
        \end{equation}
        where $\sigma_{\bar{\mat{\Gamma}},i}(f)$ denotes the $i$th singular value of $\bar{\mat{\Gamma}}(f)$, and $\zeta$ is chosen such that for $\bar{\kappa} \triangleq \eta^2 (1-\frac{2 \, \eta^2}{3 \, 2^{2b}})^{-1}$ it holds
        \begin{equation}
            \label{eq:theorem_1_eq_const}
            \frac{\bar{\kappa} \, T_\mathrm{s} }{K } \int_{-\frac{f_\mathrm{s}}{2}}^{\frac{f_\mathrm{s}}{2}} \sum_{i=1}^{\min(K,\bar{M})} \left[\bar{\mat{\Sigma}}_{\mat{H}}(f)\right]_{i,i}^2 \mathrm{d} f = 1.
        \end{equation}
        \item $\bar{\mat{U}}_{\mat{H}}(f) \in \mathbb{C}^{K \times K}$ is a unitary matrix ensuring that $\bar{\mat{U}}_{\mat{H}}(f) \bar{\mat{\Sigma}}_{\mat{H}}(f) \bar{\mat{\Sigma}}^H_{\mat{H}}(f) \bar{\mat{U}}^H_{\mat{H}}(f)$ has  identical diagonal entries. It can be obtained using \cite[Algorithm 2.2]{palomar2007mimo}.
    \end{itemize}
    Furthermore,
    the resulting minimum achievable \gls{mse} is given by\looseness-1
    \begin{IEEEeqnarray}{ll}
        \label{eq:theorem_mse_H_opt}
        & \mathbb{E} \left\lbrace \left\Vert \tilde{\matr{s}} - \hat{\matr{s}} \right\Vert^2 \right\rbrace = \E \left\lbrace \Vert \tilde{\matr{s}} \Vert^2  \right\rbrace  \nonumber \\
        & \quad - \int_{-\frac{f_\mathrm{s}}{2}}^{\frac{f_\mathrm{s}}{2}} \sum_{i=1}^{\min(K,\bar{M})} \frac{ \sigma_{\bar{\mat{\Gamma}},i}^2(f) \left[\bar{\mat{\Sigma}}_{\mat{H}}(f)\right]_{i,i}^2 }{ \left[\bar{\mat{\Sigma}}_{\mat{H}}(f)\right]_{i,i}^2 + {2^{-2b}}} \, \mathrm{d} f.
    \end{IEEEeqnarray}
\end{theorem}
\begin{IEEEproof}
    The proof is provided in Appendix \ref{sec:appendix_B}.
\end{IEEEproof}

\smallskip
Theorem \ref{th:optimal_pre_filter} characterizes the analog filter $\mat{H}^{\mathrm{o}}(f)$  in the frequency range $|f| \leq (\Upsilon+\frac{1}{2}) f_{\rm s}$. As this range coincides with the spectral support of $\matr{x}(t)$, the frequency response of the analog filter can be nullified outside this spectrum. Under the considered system model, the \gls{mse}-minimizing task-based \gls{adc} is characterized by the analog filter $\mat{H}^{\mathrm{o}}(f)$ combined with the corresponding digital filter $\mat{G}^{\mathrm{o}}[n]$, given by Proposition~\ref{th:optimal_digital_filter}.

In the following, we specialize  Theorem~\ref{th:optimal_pre_filter} to  bandlimited inputs with sampling rates satisfying the Shannon-Nyquist sampling theorem \cite[Th.~4.1]{eldar2015sampling}. The result is stated in the following corollary:
\begin{corollary}\label{cor:H_opt_Nyquist}
    For bandlimited inputs sampled at a rate satisfying  the Shannon-Nyquist sampling theorem, i.e.,
    $\left[\mat{C}_\mat{x}(f)\right]_{i,j} = 0$ for all $ |f| > \frac{f_\mathrm{s}}{2}$, $i,j \in \mathcal{M}$,  the \gls{mse}-minimizing analog filter $\mat{H}^{\mathrm{o}}(f) \in \mathbb{C}^{K \times M}$ is given by
    \begin{equation}
        \label{eq:corollary_H_opt}
        \mat{H}^{\mathrm{o}}(f) = \bar{\mat{U}}_{\mat{H}}(f) \bar{\mat{\Sigma}}_{\mat{H}}(f) \bar{\mat{V}}^H_{\mat{H}}(f) \left( \mat{C}_{\matr{x}}^{1/2}(f) \right)^{\dagger},
    \end{equation}
    where $(\cdot)^{\dagger}$ denotes the pseudo-inverse and the matrix functions $\bar{\mat{U}}_{\mat{H}}(f)$,  $\bar{\mat{\Sigma}}_{\mat{H}}(f)$, and $\bar{\mat{V}}^H_{\mat{H}}(f)$ are detailed in Theorem \ref{th:optimal_pre_filter} with $\Upsilon = 0$.\looseness-1
\end{corollary}
\begin{IEEEproof} 
    The corollary is obtained by setting $\Upsilon =0$, which corresponds to Nyquist-rate sampling, in Theorem \ref{th:optimal_pre_filter}. Then, combining \eqref{eq:def_H_bar} and \eqref{eq:theorem_H_opt} and utilizing the pseudo-inverse yields \eqref{eq:corollary_H_opt}. Note that it is necessary to utilize the pseudo-inverse here, because the \gls{psd} $\mat{C}_{\matr{x}}(f)$ might be rank deficient for some $f \in [-\frac{f_\mathrm{s}}{2},\frac{f_\mathrm{s}}{2}]$.
\end{IEEEproof}

\smallskip
\begin{figure*}[ht]
    \centering
    \includegraphics[width=\linewidth]{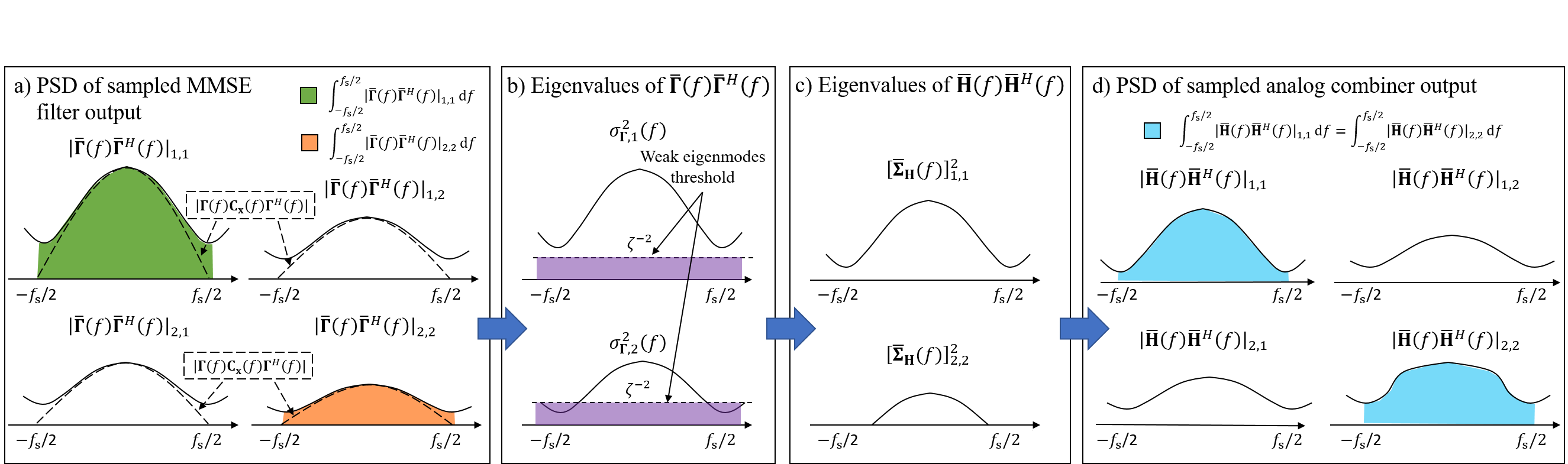}
    \caption{Illustration of the operation of the \gls{mse}-minimizing analog filter $\mat{H}^{\mathrm{o}}(f)$ for $K=N=2$.\vspace{-0.2cm}}
    \label{fig:IllustrationFlow}
    \vspace{-0.4cm}
\end{figure*}
In order to understand the operation of the \gls{mse}-minimizing analog filter detailed in Theorem~\ref{th:optimal_pre_filter}, we note that the matrix functions $\bar{\mat{\Gamma}}(f)\bar{\mat{\Gamma}}^H(f)$ and $\bar{\mat{H}}(f)\bar{\mat{H}}^H(f)$, given in \eqref{eq:def_Gamma_bar}-\eqref{eq:def_H_bar}, represent the \glspl{psd} of the outputs of the analog \gls{mmse} filter $\mat{\Gamma}(f)$ and the analog filter $\mat{H}(f)$, respectively, after being sampled at rate $f_{\rm s}$. When the analog filter is $\mat{H}^{\mathrm{o}}(f)$, given by \eqref{eq:theorem_H_opt}, two main differences are noted between $\bar{\mat{H}}(f)\bar{\mat{H}}^H(f)$ and $\bar{\mat{\Gamma}}(f)\bar{\mat{\Gamma}}^H(f)$: First, the analog combiner $\mat{H}^{\mathrm{o}}(f)$ nullifies the weak eigenmodes of  $\bar{\mat{\Gamma}}(f)$ which become indistinguishable after coarse quantization, by setting a 'water-filling'-type expression with threshold $\zeta^{-1}$. This step allows task-based \glspl{adc} to trade-off the estimation error and the distortion induced by quantization, in a manner which minimizes the overall \gls{mse} in recovering the task $\matr{s}$: Nullifying an eigenmode results in an estimation error; however, it allows to resolve the remaining eigenmodes with a higher resolution. Thereafter, $\mat{H}^{\mathrm{o}}(f)$ scales and rotates the remaining eigenmodes to minimize the maximal variance of the input signal to each quantizer, and particularly, tunes the quantizer inputs to have identical variances which exactly fit to a quantizer with unity dynamic range (cf. \eqref{eq:def_dynamic_range_squared}).

To visualize how the \gls{mse}-minimizing analog filter $\mat{H}^{\mathrm{o}}(f)$ operates, we depict in Fig. \ref{fig:IllustrationFlow} a schematic illustration comparing the \glspl{psd} of the sampled outputs of the analog \gls{mmse} filter $\mat{\Gamma}(f)$ and the analog filter $\mat{H}^{\mathrm{o}}(f)$. We visualize a setup in which $\Upsilon = 1$ and $K=N=2$, i.e., sampling is carried out below the Nyquist rate, and the number of \glspl{adc} equals the dimensionality of the task.
In particular, Fig.~\ref{fig:IllustrationFlow}.a depicts the multivariate \gls{psd} of the sampled output of the analog \gls{mmse} filter, highlighting the fact that its entries have different variances, represented by the area under $[\bar{\mat{\Gamma}}(f)\bar{\mat{\Gamma}}^H(f)]_{i,i}$.
Consequently, to uniformly quantize this signal without overloading one would have to tune the dynamic range to fit the input with the maximal variance, resulting in increased quantization distortion.
To mitigate this distortion, the weak eigenmodes are identified in Fig.~\ref{fig:IllustrationFlow}.b and used to generate the corresponding eigenvalues of $\bar{\mat{H}}(f)\bar{\mat{H}}^H(f)$ via thresholding and scaling in Fig. \ref{fig:IllustrationFlow}.c.
Finally, by applying the unitary matrix $\bar{\mat{U}}_{\mat{H}}(f)$, the resulting quantizer input in Fig. \ref{fig:IllustrationFlow}.d has identical variances which fit its unity dynamic range, thus minimizing the distortion induced by quantization.

Theorem \ref{th:optimal_pre_filter} and Corollary \ref{cor:H_opt_Nyquist} are stated for fixed \gls{adc} configuration triplets $(K, f_{\rm s}, b)$. Consequently, these results do not immediately reveal how such a triplet is to be set in light of a bit rate constraint $R$ as in \eqref{eqn:Objective}. For instance, one would be interested in understanding whether it is preferable to oversample, i.e., increase $f_{\rm s}$, while decreasing the number of bits per sample $b$.
While we do not analytically derive the  triplet which minimizes \eqref{eq:theorem_mse_H_opt}, we provide guidelines arising from Theorem \ref{th:optimal_pre_filter} in Subsection~\ref{sec:practical_design_guidelines}. Furthermore, we discuss how Theorem \ref{th:optimal_pre_filter} facilitates numerically optimizing these quantities in Subsection \ref{subsec:discussion} and provide an example in Section~\ref{sec:num_res}.

\vspace{-0.2cm}
\subsection{Practical Design Guidelines}
\label{sec:practical_design_guidelines}
\vspace{-0.1cm}
In this subsection, we provide design guidelines for practical systems that arise as a consequence of Theorem \ref{th:optimal_pre_filter}, stated in the previous subsection. In addition, we study conditions under which the intuitive design strategies of setting the analog filter to be the analog \gls{mmse} estimator or of recovering purely in the digital domain are capable of minimizing the \gls{mse}. 

The following corollary provides an upper bound on the \gls{mse}-minimizing choice of the number of \glspl{adc} $K$, when operating under a fixed rate budget:
\begin{corollary}\label{cor:optimal_K}
    Under a fixed rate budget $R = K \cdot f_\mathrm{s} \cdot b$, the \gls{mse}-minimizing number of \glspl{adc} $K$ satisfies
    \begin{equation}
        K \leq \max_{f \in \left[ -\frac{f_\mathrm{s}}{2}, \frac{f_\mathrm{s}}{2} \right]} \rank{\bar{\mat{\Gamma}}(f)}.
    \end{equation}
\end{corollary}
\begin{IEEEproof}
    The proof is provided in Appendix \ref{sec:appendix_C}.
\end{IEEEproof}

\smallskip
Corollary \ref{cor:optimal_K} implies that the number of \glspl{adc} $K$ should not be chosen larger than the maximum rank of the aliased spatial \gls{psd} of the analog \gls{mmse} filter output. This effectively implies that using a larger number of \glspl{adc} than that required to represent the sampled output of the analog \gls{mmse} filter results in unnecessary distortion. In such cases, one can use fewer \glspl{adc} with higher resolution, i.e., larger $b$, without losing sufficiency \wrt the \gls{mmse} estimate $\tilde{\matr{s}}$. This is achieved by utilizing  the \gls{mse}-minimizing analog filter detailed in Theorem~\ref{th:optimal_pre_filter} to further trade sufficiency for quantization distortion in a manner which minimizes the overall \gls{mse}. In particular, Corollary \ref{cor:optimal_K} implies that one should use $K\leq N$ \glspl{adc}, but does not specify the \gls{mse}-minimizing value. In some scenarios the \gls{mse} is optimized by using $K$ smaller than $N$, while in other cases it is preferable to set $K=N$, as we numerically demonstrate in Section~\ref{sec:num_res}.

An additional insight which can be obtained analytically is that increasing the quantization resolution improves the \gls{mse} more notably compared to oversampling. This is obtained based on the following corollary, which investigates the trade-off between $f_\mathrm{s}$ and $b$, when operating under a fixed rate budget and sampling above the Nyquist rate, denoted as $f_\mathrm{nyq}$:
\begin{corollary}\label{cor:trade-off_fs_b}
    Under a fixed rate budget $R = K \cdot f_\mathrm{s} \cdot b$, for a fixed $K$ and for $f_\mathrm{s} \geq f_\mathrm{nyq}$, the \gls{mse} is minimized by maximizing $b$.
\end{corollary}
\begin{IEEEproof}
    The proof is provided in Appendix \ref{sec:appendix_C}.
\end{IEEEproof}

\smallskip
 Corollary \ref{cor:trade-off_fs_b} implies that for a given number of \glspl{adc} and a fixed bit rate allowing Nyquist rate sampling,  Nyquist rate sampling with increased quantization resolution allows achieving a lower \gls{mse} compared to oversampling, \iec choosing $f_\mathrm{s} > f_\mathrm{nyq}$, with lower resolution.
However, it is emphasized that the considered system model does not allow for feedback from the digital to the analog part, and thus does not cover Sigma-Delta architectures \cite{aziz1996overview}, which are known to benefit from oversampling.

Next, we compare the  task-based acquisition system derived in Subsection \ref{sec:main_results} to two intuitive alternative designs:
\begin{itemize}
    \item \emph{Analog recovery}, where the task is estimated solely in the analog domain, i.e., $\mat{H}(f)$ is set to be the analog \gls{mmse} filter $\mat{\Gamma}(f)$, and its output is subsequently converted into a digital representation using $K=N$ \glspl{adc}.
    \item \emph{Digital recovery}, where no analog processing is employed, i.e., $\mat{H}(f) \equiv \mat{I}_M$. In such cases, the \gls{ct} random process $\matr{x}(t)$ is digitized by $K=M$ \glspl{adc}, and the task is subsequently estimated fully in the digital domain.
\end{itemize}
These alternative designs can be understood as the two possible extreme cases of carrying out the estimation of the task solely in the analog or digital domain. As Theorem \ref{th:optimal_pre_filter} implies that in general $\mat{H}^{\mathrm{o}}(f) \neq \mat{\Gamma}(f)$ and $\mat{H}^{\mathrm{o}}(f) \neq \mat{I}_M$, it can be concluded that none of the two alternative designs is generally optimal in minimizing the \gls{mse}. Corollary \ref{cor:optimal_K} indicates that a digital-only strategy can only minimize the \gls{mse} when the cardinality of the task vector $N$ is at least as large as the number of analog signals $M$. When this is not the case, it leads to degraded performance compared to our task-based design. As for analog recovery, in the following corollary we identify conditions for this strategy to minimize the \gls{mse} in recovering $\matr{s}$:
\begin{corollary}\label{cor:analog_recovery_opt}
    Analog recovery, \iec $\mat{H}^\mathrm{o}(f) = \mat{\Gamma}(f)$ and hence $K=N$, minimizes the \gls{mse} if and only if the analog \gls{mmse} estimate $\tilde{\matr{s}}$ consists of uncorrelated entries with identical variance, i.e., if its covariance matrix $\mat{C}_{\tilde{\matr{s}}}$ is a scaled identity matrix, i.e., $\mat{C}_{\tilde{\matr{s}}} = c \, \mat{I}_N$, for some $c \geq 0$.
\end{corollary}
\begin{IEEEproof}
    The proof is provided in Appendix \ref{sec:appendix_D}.
\end{IEEEproof}

\vspace{-0.2cm}
\subsection{Asymptotic Analysis}\label{sec:asymptotic_analysis}
\vspace{-0.1cm}
In this subsection, we investigate the \gls{mse} behavior for an asymptotically large bit rate $R$. Clearly, by Corollary \ref{cor:optimal_K}, using an arbitrarily large number of \glspl{adc} $K$ is not optimal. Therefore, in the following we focus on the remaining parameters in the triplet $(K, f_{\rm s}, b)$, and characterize the \gls{mse} of the proposed task-based \gls{adc} system with infinitesimally dense sampling, \iec for $T_\mathrm{s} \rightarrow 0$, and with an infinite quantization resolution, \iec for $b \rightarrow \infty$.
While this asymptotic analysis no longer represents practical systems operating with finite sampling rate and quantization resolution, it reveals the individual effect of each of the two \gls{adc} operations, i.e., sampling and quantization, on the accuracy of recovering the task $\matr{s}$. 

The following analysis is carried out focusing on bandlimited signals (though possibly with a very large bandwidth), as done in Theorem~\ref{th:optimal_pre_filter}, and holds for \emph{any} analog filter $\mat{H}(f)$ which satisfies the following condition:
\begin{equation}
    \label{eq:condition_H_maintains_sufficiency}
    \bar{\mat{H}}^H(f) \left( \bar{\mat{H}}(f) \bar{\mat{H}}^H(f) \right)^{-1} \bar{\mat{H}}(f) \bar{\mat{\Gamma}}^H(f) = \bar{\mat{\Gamma}}^H(f),
\end{equation}
for all $f \in \big[ -\frac{f_\mathrm{s}}{2}, \frac{f_\mathrm{s}}{2}\big]$. The \gls{lhs} of \eqref{eq:condition_H_maintains_sufficiency} is the projection of $\bar{\mat{\Gamma}}^H(f)$ onto the column space of $\bar{\mat{H}}^H(f)$, \iec \eqref{eq:condition_H_maintains_sufficiency} holds if $\bar{\mat{\Gamma}}^H(f)$ lies in the column space of $\bar{\mat{H}}^H(f)$ for all $f \in \big[ -\frac{f_\mathrm{s}}{2}, \frac{f_\mathrm{s}}{2}\big]$. In this case, we say that the analog filter $\mat{H}(f)$ maintains sufficiency \wrt the analog \gls{mmse} estimate $\tilde{\matr{s}}$.
Note that if $\bar{\mat{\Sigma}}_{\mat{H}}(f)$ contains $K$ non-zero diagonal entries, then the condition in \eqref{eq:condition_H_maintains_sufficiency} is satisfied by the \gls{mse}-minimizing analog filter $\mat{H}^\mathrm{o}(f)$ (cf.~\eqref{eq:theorem_H_opt}).

First, we characterize the achievable \gls{mse} expression of Proposition~\ref{th:optimal_digital_filter}  for $T_\mathrm{s} \rightarrow 0$, i.e., for infinitesimally dense sampling, as stated in the following lemma:
\begin{lemma}\label{lem:asym_MSE_Ts}
    For any analog filter $\mat{H}(f)$ satisfying \eqref{eq:condition_H_maintains_sufficiency}, the achievable \gls{mse} tends to zero as $T_\mathrm{s} \rightarrow 0$.
\end{lemma}

\begin{IEEEproof}
The proof is provided in Appendix \ref{sec:appendix_E}.
\end{IEEEproof}

\smallskip
Lemma \ref{lem:asym_MSE_Ts} shows that the  \gls{mse} vanishes for infinitesimally dense sampling. In fact, it indicates that even for 1-bit quantization, the distortion vanishes with infinite oversampling, as the digital processor has an arbitrarily large number of samples from which it can recover the finite-dimensional $\matr{s}$, and the \gls{mse} tends to zero when using dithered quantizers.

Next, we investigate the  \gls{mse}  of Proposition~\ref{th:optimal_digital_filter} for $b \rightarrow \infty$, i.e., for an infinite number of quantization levels. This asymptotic \gls{mse} is stated in the following lemma: 
\begin{lemma}\label{lem:asym_MSE_b}
    For any analog filter $\mat{H}(f)$ satisfying \eqref{eq:condition_H_maintains_sufficiency}, the achievable \gls{mse} tends to zero as $b \rightarrow \infty$.
\end{lemma}

\begin{IEEEproof}
The proof is provided in Appendix \ref{sec:appendix_E}.
\end{IEEEproof}

\smallskip
Lemma \ref{lem:asym_MSE_b} shows that even for sub-Nyquist sampling, the estimation error tends to zero for an asymptotically large number of quantization levels, if the analog filter $\mat{H}(f)$ maintains sufficiency \wrt the task. This holds, since the \gls{mmse} estimate is only the sample taken at time instance $t=0$ at the output of the analog \gls{mmse} filter $\mat{\Gamma}(f)$, and thus can be recovered from its sampled output regardless of the sampling rate.
Note that for $b \rightarrow \infty$, the resulting problem can be considered as \emph{sampling for a task}.
Here, the fact that by Lemma \ref{lem:asym_MSE_b} one can achieve arbitrarily small \gls{mse} indicates that one can substantially benefit in terms of sampling rate when accounting for the presence of a task. However, it is noted that the ability to approach zero \gls{mse} could follow from the fact that the task is a vector that has no bandwidth. Hence, it is unclear if a similar result holds when one seeks to recover a process rather than a vector.\looseness-10

While Lemmas \ref{lem:asym_MSE_Ts} and \ref{lem:asym_MSE_b} imply that increasing both the sampling rate and the quantization resolution can result in arbitrarily small \gls{mse}, our numerical study in Section~\ref{sec:num_res} demonstrates that increasing the quantization resolution typically results in more substantial \gls{mse} reduction compared to increasing the sampling rate. 

\vspace{-0.2cm}
\subsection{Discussion}
\label{subsec:discussion}
\vspace{-0.1cm}
The task-based \glspl{adc}, as derived in the previous subsections, combine the designs proposed for task-based \emph{quantization} systems in \cite{shlezinger2018hardwarelimited}, which did not consider sampling, with multivariate sampling concepts, as studied in \cite{shlezinger2019joint}.
In particular, we show that by additionally jointly accounting for the presence of uniform samplers in the overall acquisition system design, one can mitigate the error due to quantization by employing oversampling (cf.~Lemma~\ref{lem:asym_MSE_Ts}), or alternatively utilize sub-Nyquist sampling to assign more bits, thus possibly reducing the \gls{mse} under tight rate budgets, as we numerically show in Subsection~\ref{sec:num_res_rate_budget}. Comparing to the previous work \cite{8350400}, we note that this work studied the recovery of processes via joint sampling and lossy source coding and is thus fundamentally different from our work. Nonetheless, one can identify similar insights which arise from our task-based hardware-limited design and \cite{8350400}, \iec that sub-Nyquist sampling allows us to reduce the \gls{mse} further under low bit rate budgets. We consider extending the framework studied here to recover task processes rather than vectors as a possible future research direction.\looseness-1

As detailed in the problem formulation in Section \ref{sec:system_model}, we limit our analysis of task-based \glspl{adc} to the class of \emph{linear} tasks. Nevertheless, an extension to larger classes, i.e., non-linear tasks, is possible based on the proposed scheme. For instance, task-based quantization for the class of \emph{quadratic tasks} was studied in \cite{salamatian2019task} by utilizing principal inertia component methods to convert the task function into a linear one. A similar approach could be used to extend the concept of task-based \glspl{adc}, which is proposed in this work, to non-linear tasks, and we leave this study for future work. 
Furthermore, in practical hardware implementations of hybrid acquisition systems, one must account for additional noise sources and impairments which are not considered in our system model. While some of these noise sources can be incorporated into the dither signal, their modeling is highly implementation-dependent. We thus leave the extension of our results to hardware implementation with additional noise sources for future research.

The equivalent quantizer model detailed in Subsection~\ref{sec:quantization} models the quantization error as an additive distortion, which is uncorrelated to the quantizer input. This model holds rigorously for the considered overload-free nonsubtractive dithered quantizers. However, it also holds \emph{approximately} for conventional (non-dithered) uniform quantizers for a broad range of inputs, and notably, for sub-Gaussian input signals \cite[Sec.~VIII]{widrow1996statistical}.
The critical conditions for this to hold approximately are a negligible overload probability, and a sufficiently large input variance compared to the quantization step size.
Hence, even though we assume nonsubtractive dithered quantizers, the results can also be applied to systems employing conventional quantizers, as numerically verified in Subsection~\ref{sec:num_res_eval_framework}.
Similar result have been obtained in \cite[Sec.~VI]{shlezinger2018hardwarelimited}.

Our work considers the general case of jointly \gls{wss} input signals $\left\lbrace \mathsf{x}_m(t) \right\rbrace_{m=1}^{M}$ and utilizes the structure, which is given by the task $\matr{s}$, to optimize the task-based \gls{adc} system. However, depending on the application and the specific task, the input signals might exhibit additional structure, which can be further exploited, as usually done in sub-Nyquist sampling \cite{eldar2015sampling,5419072}. This additional structure could be exploited to either further reduce the \gls{mse} or to reduce the hardware complexity, i.e., reducing $K$, $f_\mathrm{s}$, or $b$, to achieve the same \gls{mse}.

In Subsection~\ref{sec:main_results}, the \gls{mse}-minimizing task-based acquisition system is characterized for a fixed \gls{adc} configuration $(K, f_\mathrm{s}, b)$ by combining Proposition~\ref{th:optimal_digital_filter} and Theorem~\ref{th:optimal_pre_filter}. However, our actual goal is to obtain the \gls{mse}-minimizing task-based acquisition system under a rate budget $R \geq K \cdot f_\mathrm{s} \cdot b$. Utilizing the results from Proposition~\ref{th:optimal_digital_filter} and Theorem~\ref{th:optimal_pre_filter}, the \gls{mse}-minimizing \gls{adc} configuration $(K^\mathrm{o}, f_\mathrm{s}^\mathrm{o}, b^\mathrm{o})$ for a given task can be found numerically: $K$ and $b$ are discrete variables and their choice is typically upper-bounded in practice. For example Corollary~\ref{cor:optimal_K} implies that the search over $K$ can be restricted to $K \leq \max \rank{\bar{\mat{\Gamma}}(f)}$. Hence, a grid search over all feasible combinations of $K$ and $b$ can be employed. Furthermore, because the \gls{mse} is minimized by fully utilizing the available bit rate budget $R$, \iec when $R = K \cdot f_\mathrm{s} \cdot b$, we can choose $f_\mathrm{s} = \frac{R}{K \cdot b}$. Note that the triplet $(K^\mathrm{o}, f_\mathrm{s}^\mathrm{o}, b^\mathrm{o})$ is determined in the process of system design, i.e., the optimization can be performed offline. If grid search is computationally prohibitive, \egc due to a large range of $K$ and $b$, one can resort to more efficient optimization techniques such as random search methods or Bayesian optimization \cite{brochu2010tutorial}. We provide an example for such a numerical evaluation using a grid search in Subsection~\ref{sec:num_res_rate_budget}.

The task-based \gls{adc} system considered in this work employs both analog and digital signal processing in a hybrid manner. In Subsection~\ref{sec:main_results} we obtain the \gls{mse}-minimizing analog and digital filters, \iec the optimal split between analog and digital processing, under a fixed \gls{adc} configuration $(K, f_\mathrm{s}, b)$. This results can then be utilized to find the \gls{mse}-minimizing parameters of the \glspl{adc}, \iec $(K^\mathrm{o}, f_\mathrm{s}^\mathrm{o}, b^\mathrm{o})$, under a budget on the digital rate $R \geq K \cdot f_\mathrm{s} \cdot b$, as discussed before. While the rate budget $R$ also relates to the power consumption and implementation complexity of the \glspl{adc}, it does not take into account the additional complexity of the analog processing. Furthermore, in some applications, the filters of which the hybrid system is comprised may be constrained to some structure. For instance, one may be interested in focusing on causal finite-length digital filters, while the analog filter may be required to represent an interconnection of phase shifters and adders \cite{mendez2016hybrid}, or be subject to restrictions following from the antenna design \cite{wang2019dynamic}. These restrictions, as well as additional design constraints which arise from the specific application, are likely to reflect on the cost and the implementation complexity of the system.
The design of the system subject to these additional considerations is beyond the scope of the current work.\looseness-1

\vspace{-0.2cm}
\section{Numerical Results}\label{sec:num_res}
\vspace{-0.1cm}
Here, we evaluate the  proposed task-based \gls{adc} system in terms of its achievable \emph{normalized \gls{mse}}, given by $\frac{\mathbb{E} \left\lbrace \left\Vert \tilde{\matr{s}} - \hat{\matr{s}} \right\Vert^2 \right\rbrace}{\E \left\lbrace \Vert \tilde{\matr{s}} \Vert^2  \right\rbrace}$. We compare the performance of the proposed task-based acquisition system, i.e., $\mat{H}(f)=\mat{H}^\mathrm{o}(f)$, to the two alternative design strategies of analog recovery, i.e., $\mat{H}(f)=\mat{\Gamma}(f)$, and digital recovery, i.e., $\mat{H}(f)=\mat{I}_M$. We use   Theorem~\ref{th:optimal_pre_filter},  to evaluate the \gls{mse} of the proposed system, and  Proposition~\ref{th:optimal_digital_filter} to evaluate the \gls{mse} for analog and digital recovery, respectively.

\vspace{-0.2cm}
\subsection{MIMO Matched Filtering}
\label{sec:num_res:mimo_matched_filtering}
\vspace{-0.1cm}
As an example, we consider the problem of estimating the output of a set of matched filters.
Matched filtering is an essential and standard operation in processing analog signals acquired by communications receivers \cite[Ch. 5.1]{goldsmith2005wireless}.
In digital communications systems, matched filters are usually applied in the digital domain \cite[Sec.~4.3.3]{meyr1997}.
Namely, the signals are first converted into a digital representation in a generic task-ignorant manner, after which they are filtered in digital.
However, this approach is expected to be sub-optimal for systems with bit-limited \glspl{adc}, and, thus, in the following numerical example, we evaluate the gains one can achieve by accounting for this task during the signal acquisition by using task-based \glspl{adc}.
Since complex-valued channels can be equivalently represented as multivariate real-valued channels of extended dimensions, we consider a multivariate real-valued model, which matches the signal model detailed in Subsection~\ref{sec:Signal}.\looseness-1

Let $\matr{x}(t)$ be the multivariate observations at a \gls{mimo} receiver,  given by
\begin{equation}
    \matr{x}(t) = \left( \mat{F} * \tilde{\matr{x}} \right)(t) + \matr{n}(t),
    \label{eq:num_res_x_def}
\end{equation}
where $\tilde{\matr{x}}(t)$ denotes a Gaussian transmit signal with autocorrelation function $\mat{C}_{\tilde{\matr{x}}}(\tau) = \E\{ \tilde{\matr{x}}(t+\tau) \tilde{\matr{x}}^T(t) \} = \mat{I}_{N} \delta(\tau)$, $\mat{F}(t) = \tilde{\mat{F}} \, \frac{1}{T_\mathrm{nyq}}\mathrm{sinc}(\frac{t}{T_\mathrm{nyq}})$, $ \tilde{\mat{F}} \in \mathbb{R}^{M \times N}$ denotes the channel, and $\matr{n}(t)$ is an additive white Gaussian noise process with autocorrelation function $\mat{C}_{\matr{n}}(\tau) = \E\{ \matr{n}(t+\tau) \matr{n}^T(t) \} = \frac{N_0}{2} \mat{I}_{M} \frac{1}{T_\mathrm{nyq}}\mathrm{sinc}(\frac{\tau}{T_\mathrm{nyq}})$. Note that the channel $\mat{F}(t)$ and noise $\matr{n}(t)$ are bandlimited to $f \in [-\frac{f_\mathrm{nyq}}{2},\frac{f_\mathrm{nyq}}{2}]$ with $f_\mathrm{nyq} = \frac{1}{T_\mathrm{nyq}}$. The task is to estimate the noiseless matched filter output, i.e.,
\begin{equation}
    \matr{s} = \left( \cev{\mat{F}}^T * \mat{F} * \tilde{\matr{x}} \right)(0).
    \label{eq:num_res_task_def}
\end{equation}
From the orthogonality principle \cite{papoulis2001probability}, it follows that the analog \gls{mmse} estimator $\mat{\Gamma}(f)$ is given by $\mat{\Gamma}(f) = \mat{C}_{\matr{s}\matr{x}}(f) \mat{C}^{-1}_{\matr{x}}(f)$.

With the above, we evaluate the performance for $N=4$ and $M=16$, i.e., $4$ transmit and $16$ receive antennas. The filter bandwidth is chosen as $f_\mathrm{nyq}=\SI{400}{\mega \hertz}$, which corresponds to the largest channel bandwidth of the 5G NR \gls{mmwave} bands \cite{3gpp.38.101}.
The channel is modeled as $\tilde{\mat{F}} = \mat{C}_{\mathrm{Rx}}^{1/2} \mat{H}_{\mathrm{ch}}$, where $\mat{C}_{\mathrm{Rx}} \in \mathbb{R}^{M \times M}$ models the spatial correlation at a uniform linear array receiver for impinging received signals with angular spread $\sigma_\phi$. $\mat{C}_{\mathrm{Rx}} $ is given by (cf.~ \cite[eq.~(19)]{4277071})
\begin{equation}
    \left[\mat{C}_{\mathrm{Rx}}\right]_{m,n} = \frac{ \left(1-e^{-\sqrt{2} \pi / \sigma_\phi} \right)^{-1} }{ 1 + \frac{\sigma_\phi^2}{2}(\pi(m - n))^2 }.
    \label{eq:C_Rx}
\end{equation}
Because \eqref{eq:C_Rx} is accurate for small angular spreads \cite{4277071}, we choose $\sigma_\phi = \SI{1}{\degree}$.
Moreover, $\mat{H}_{\mathrm{ch}} \in \mathbb{R}^{M \times N}$ is assumed to be known and fixed.
It contains independent entries which are generated randomly as $\left[\mat{H}_{\mathrm{ch}}\right]_{m,n} \sim \mathcal{N}(0,1)$\footnote{\label{ftn:reproduce} Reproducible in MATLAB R2021a by setting: \texttt{rng(0), H\_ch = randn(16,4)}.}.
Furthermore, we define $\mathrm{SNR} \triangleq \frac{2}{N_\mathrm{0}} \trace{ \tilde{\mat{F}} \tilde{\mat{F}}^T} = \SI{10}{\decibel}$.

\vspace{-0.2cm}
\subsection{Evaluation of the Theoretical Results}\label{sec:num_res_eval_framework}
\begin{figure}
	\centering
	\includegraphics[width=\columnwidth]{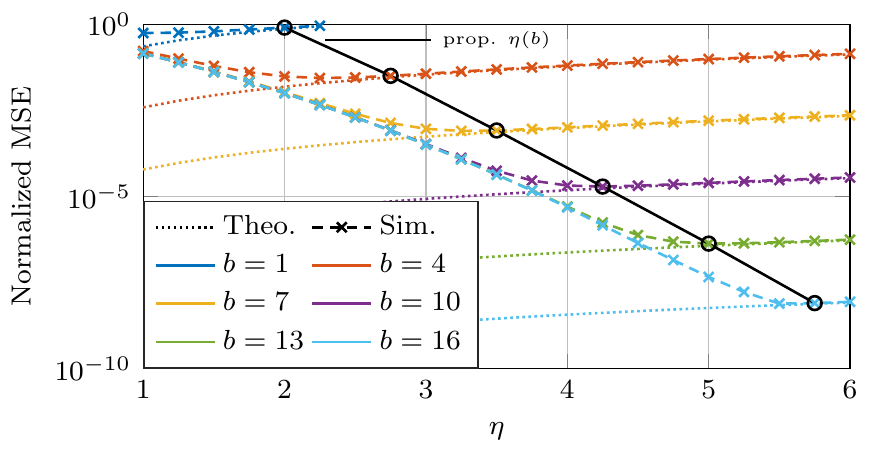}
	\vspace{-0.7cm}
	\caption{Normalized \gls{mse} vs. $\eta$ for $N=M=K=1$ and $f_\mathrm{s}=f_\mathrm{nyq}$ when employing dithered quantizers with different resolutions $b$. For $\eta(b)=0.25b+1.75$ (depicted as a black solid line), the simulated \gls{mse} is approx. \SI{5}{\percent} larger than the \gls{mse} given in \eqref{eq:theorem_mse_H_opt}.\looseness-10 \vspace{-0.4cm}}
	\label{fig:eval_mse_over_eta}
\end{figure}

\vspace{-0.1cm}
In the following, we evaluate the accuracy of the derived \gls{mse} expressions for non-zero overload probabilities, and investigate the performance when employing conventional non-dithered quantizers. First, we compare the \gls{mse} expression given in \eqref{eq:theorem_mse_H_opt}, which is derived assuming a zero overload probability, to the simulated \gls{mse} with non-zero overload probabilities when employing dithered quantizers.
For this evaluation, we focus on the scalar case, \iec we consider $N=M=K=1$, and employ Nyquist rate sampling, \iec $f_\mathrm{s} = f_\mathrm{nyq}$.
Recalling that the overload probability in our system model is controlled by setting the dynamic range $\gamma$ to a multiple $\eta$ of the standard deviation of the quantizer inputs  (cf.~\eqref{eq:def_dynamic_range_squared}), we evaluate the normalized \gls{mse} over $\eta$ for different amplitude resolutions $b$ in Fig.~\ref{fig:eval_mse_over_eta}.
We observe that the simulated and theoretical \gls{mse} converge for sufficiently high $\eta$.
However, we also note that higher amplitude resolutions $b$ require a larger $\eta$, \iec a lower overload probability, such that both \glspl{mse} converge.
Based on the previous findings, we choose $\eta$ as a function of $b$, \iec we choose $\eta(b) = 0.25b + 1.75$, in the remainder of this work.
For this choice, the simulated \gls{mse} is approximately \SI{5}{\percent} higher than the by \eqref{eq:theorem_mse_H_opt} predicted \gls{mse}, demonstrating a good match between the simulated results, which utilize actual \gls{adc} mappings with non-zero overload probability, and the theoretical results derived based on a statistical model for the \gls{adc} operation that holds for non-overloaded quantizers.

\begin{figure}
	\centering
	\includegraphics[width=\columnwidth]{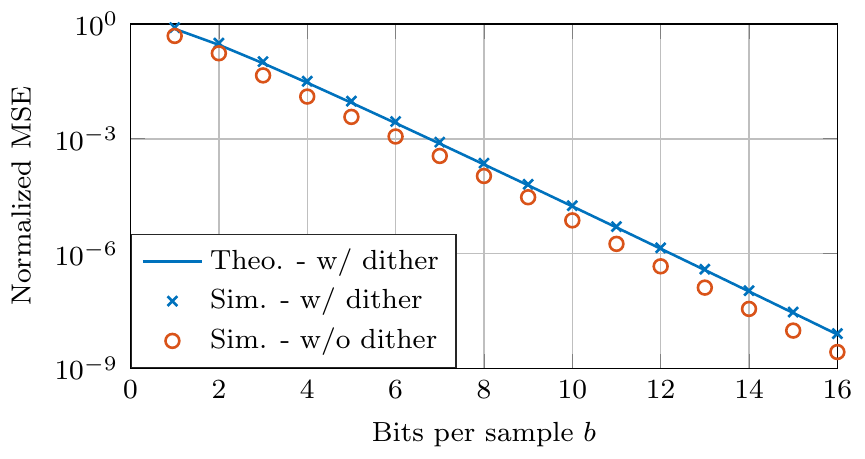}
	\vspace{-0.7cm}
	\caption{Normalized \gls{mse} vs. number of bits per sample $b$ for $N=M=K=1$ and $f_\mathrm{s}=f_\mathrm{nyq}$ when employing nonsubtractive dithered and conventional non-dithered quantizers.\looseness-10 \vspace{-0.4cm}}
	\label{fig:eval_dither}
\end{figure}

Next, we compare the \gls{mse} when employing conventional, \iec non-dithered, quantizers instead of nonsubtractive dithered quantizers, as assumed in Section~\ref{sec:system_model}, for $N=M=K=1$ and $f_\mathrm{s} = f_\mathrm{nyq}$ in Fig.~\ref{fig:eval_dither}. We observe that the theoretical \gls{mse} computed via Theorem~\ref{th:optimal_pre_filter} matches the simulated results for the dithered acquisition system. Furthermore, the normalized \gls{mse} when employing non-dithered quantizers is found to be \emph{lower} than the \gls{mse} when employing nonsubtractive dithered quantizers.
This validates the quantizer model described in Subsection~\ref{sec:quantization}.
Note that a similar result has been obtained in \cite[Sec.~VI]{shlezinger2018hardwarelimited}.
For high quantizer resolutions $b$, the gain of employing non-dithered quantizers corresponds to approx. one bit.
In the following, we continue to evaluate the performance assuming nonsubtractive dithered quantizers with zero overload probability.
Next, we evaluate the normalized \gls{mse}, when varying each of the \gls{adc} parameters $K$, $b$ and $f_\mathrm{s}$ individually, while fixing the remaining parameters.\looseness-1

\vspace{-0.2cm}
\subsection{Evaluation of the Choice of ADC Parameters}\label{sec:num_res_eval_adc_parameters}
\begin{figure}
	\centering
	\includegraphics[width=\columnwidth]{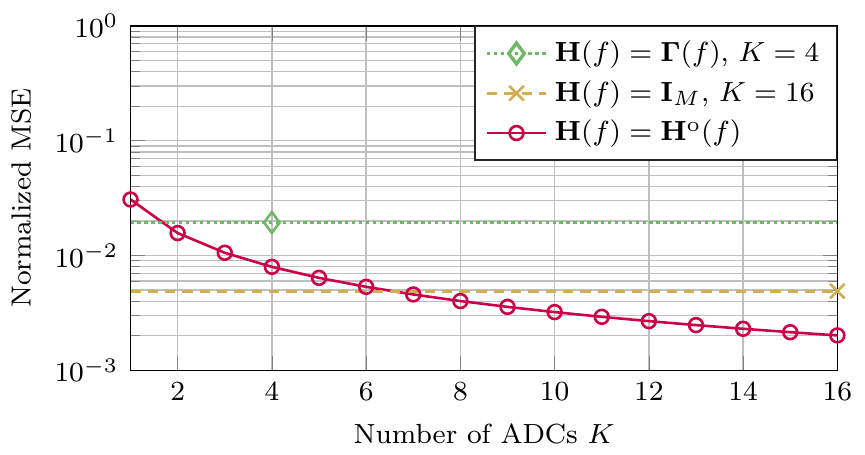}
	\vspace{-0.7cm}
	\caption{Normalized \gls{mse} vs. number of \glspl{adc} $K$,  $b=4$ and $f_\mathrm{s}=f_\mathrm{nyq}$.\looseness-10 \vspace{-0.4cm}}
	\label{fig:eval_K_journal}
\end{figure}

\vspace{-0.1cm}
First, in Fig.~\ref{fig:eval_K_journal}, we compare the normalized \gls{mse}, when varying the number of \glspl{adc} $K$ for a fixed number of bits per sample, chosen as $b=4$, and a fixed sampling rate of $f_\mathrm{s}=f_\mathrm{nyq}$.
The normalized \gls{mse} decreases monotonically with $K$.
For $K \geq 7$, the proposed task-based acquisition system outperforms digital recovery, which employs $K=16$ \glspl{adc}. Therefore, the proposed system achieves a lower \gls{mse} with \SI{56}{\percent} fewer \glspl{adc}. Note that this result does not contradict the result from Corollary~\ref{cor:optimal_K}, which states that under a constraint on the bit rate it is inefficient to choose $K$ larger than the maximum rank of the aliased analog \gls{mmse} filter $\bar{\mat{\Gamma}}(f)$.

\begin{figure}
	\centering
	\includegraphics[width=\columnwidth]{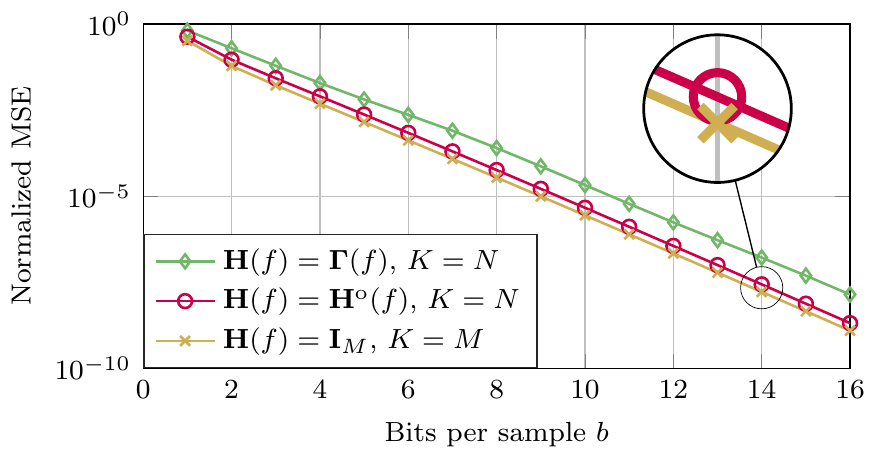}
	\vspace{-0.7cm}
	\caption{Normalized \gls{mse} vs. number of bits per sample $b$,  $f_\mathrm{s}=f_\mathrm{nyq}$. The task-based  $\mat{H}(f)=\mat{H}^\mathrm{o}(f)$ achieves nearly the same performance as digital recovery, i.e., $\mat{H}(f)=\mat{I}_M$, using  only $K=4$ instead of $K=16$ \glspl{adc}.\vspace{-0.4cm}}
	\label{fig:eval_b_journal}
\end{figure}

Next, in Fig.~\ref{fig:eval_b_journal} we evaluate the normalized \gls{mse} for Nyquist rate sampling, i.e., $f_\mathrm{s} = f_\mathrm{nyq}$, over the number of bits per sample $b$. For this evaluation, we set the number of \glspl{adc} of the proposed system to $K=N=4$ (cf.~Corollary~\ref{cor:optimal_K}). From Fig.~\ref{fig:eval_b_journal} we note that digital recovery, which utilizes $K=16$ \glspl{adc} with $b$ bits each, yields the lowest \gls{mse}. The proposed task-based \gls{adc} system, which uses only $K=4$ \glspl{adc} of the same resolution, namely, \SI{75}{\percent} fewer bits, yields a marginally higher \gls{mse}, while analog recovery results in the highest \gls{mse}. This indicates that substantial resource savings can be achieved by task-based acquisition with a marginal degradation of the overall accuracy in recovering the task vector.

\begin{figure}
	\centering
	\includegraphics[width=\columnwidth]{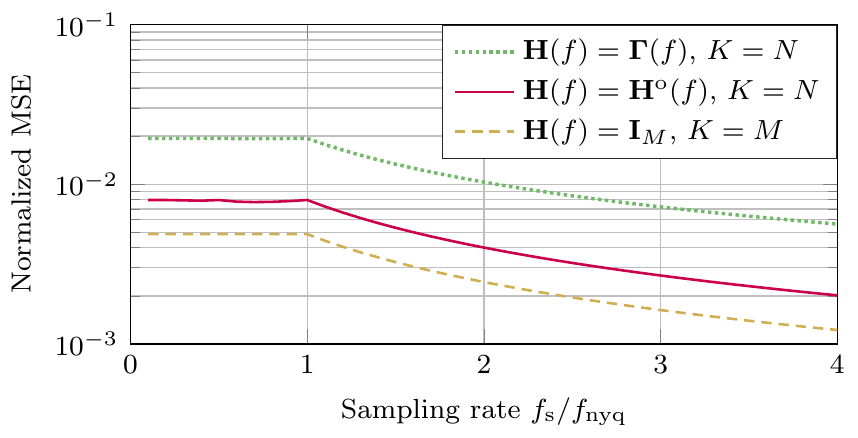}
	\vspace{-0.7cm}
	\caption{Normalized \gls{mse} vs. sampling rate $f_\mathrm{s}$,  $b=4$. For $f_\mathrm{s}>f_\mathrm{nyq}$ the \gls{mse} is decreasing monotonically with increasing $f_\mathrm{s}$. However, for $f_\mathrm{s}<f_\mathrm{nyq}$, the \gls{mse} remains approx. constant, which is discussed in Sec.~\ref{sec:num_res_eval_adc_parameters}.\vspace{-0.3cm}}
	\label{fig:eval_fs_journal}
\end{figure}

Finally, the performance when varying the sampling rate $f_\mathrm{s}$ is evaluated in Fig.~\ref{fig:eval_fs_journal} for $b=4$. For sampling rates above the Nyquist rate, i.e., for $f_\mathrm{s} > f_\mathrm{nyq}$, the \gls{mse} is monotonically decreasing with increasing $f_\mathrm{s}$, for all considered architectures. This is expected, as additional samples can be utilized to reduce the \gls{mse}. For sub-Nyquist sampling, i.e., for $f_\mathrm{s} < f_\mathrm{nyq}$, the \gls{mse} remains approx. constant. This result is caused by the fact that we define our task as the matched filter output at $t=0$ (cf.~\eqref{eq:num_res_task_def}), i.e., our task has no bandwidth, and we employ sub-Nyquist sampling \wrt the filtered input $\matr{y}(t)$ and not \wrt the task. Furthermore, the sampling time instances are given by $n T_\mathrm{s}$, $n \in \mathbb{Z}$. Hence, for any $f_\mathrm{s}$, there is always a sample at $t = n T_\mathrm{s} = 0$, which corresponds to a noisy observation of the task.\looseness-10

\begin{figure}
	\centering
	\includegraphics[width=\columnwidth]{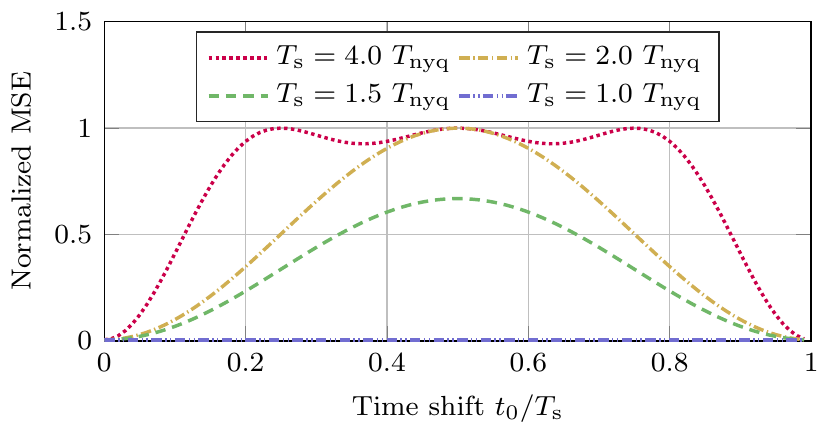}
	\vspace{-0.7cm}
	\caption{Normalized \gls{mse} for recovering $\tilde{\matr{s}}_{t_0}=(\mat{\Gamma} * \matr{x})(t_0)$ using $\mat{H}(f)=\mat{H}^\mathrm{o}(f)$ which is designed to recover $\tilde{\matr{s}}=(\mat{\Gamma} * \matr{x})(0)$ with $K=N$ and $b=4$.\vspace{-0.3cm}}
	\label{fig:eval_mse_over_t0}
\end{figure}

In order to avoid making conclusions that do not necessarily reflect matched filtering tasks, as considered here, we recall that acquisition systems are rarely designed to recover a single sample. A more likely mode of operation involves periodically applying a matched filter without modifying the analog filter for each sample. Therefore, next, we evaluate how the sampling rate affects the ability of task-based \glspl{adc} to apply matched filtering at multiple time instances, which are not necessarily known at the design stage, using the same analog hardware.
Now, our results from Subsection \ref{sec:main_results} can be used to design task-based \glspl{adc} for a specific time instant.
Neglecting the quantization error, for $f_\mathrm{s} > f_\mathrm{nyq}$ it is possible to digitally interpolate the task, \iec the matched filter output, for any time instant $t_0 \in \mathbb{R}$, i.e., $\tilde{\matr{s}}_{t_0} \triangleq (\mat{\Gamma} * \matr{x})(t_0)$ can be recovered digitally.
Consequently, designing the analog filter, \iec $\mat{H}^{\mathrm{o}}(f)$, for $t_0=0$ is not a limitation.
However, when employing sub-Nyquist sampling \wrt $\{\mathsf{y}_k(t) \}_{k=1}^K$, perfect interpolation is no longer possible and the \gls{mse} becomes periodic with $T_\mathrm{s}$ in the time shift $t_0$, as illustrated in Fig.~\ref{fig:eval_mse_over_t0}. It can be observed that for sub-Nyquist sampling the \gls{mse} grows significantly when the desired time instant $t_0$ is not close to the sampling time instances.
While Fig.~\ref{fig:eval_fs_journal} implies that when seeking to capture a quantity that can be represented as a single sample, one can utilize arbitrarily small sampling rates, sampling below the Nyquist rate limits the flexibility of the resulting system and can induce substantial losses in the presence of time shifts. The numerical results here are obtained by inserting $\tilde{\mat{\Gamma}}(f) \triangleq \mathcal{F} \left\lbrace \mat{\Gamma}(t-t_0) \right \rbrace = \mat{\Gamma}(f) e^{-j 2 \pi f t_0}$ for $\mat{\Gamma}(f)$ in \eqref{eq:def_S_of_f} in Proposition~\ref{th:optimal_digital_filter}, where $\mathcal{F} \left\lbrace \cdot \right\rbrace$ denotes the \gls{ft}.\looseness-10

\vspace{-0.2cm}
\subsection{Evaluation Under Fixed Rate Budget}\label{sec:num_res_rate_budget}
\vspace{-0.1cm}
In almost any practical application we aim to estimate the task vector at multiple time instances and it is practical to assume that the time instant can be chosen digitally. Therefore, following our observations in Figs. \ref{fig:eval_fs_journal}-\ref{fig:eval_mse_over_t0}, in order to faithfully compare the performance of task-based \glspl{adc} with different configurations, we numerically evaluate the \emph{time-averaged} \gls{mse} $\frac{1}{T_\mathrm{s}} \int_0^{T_\mathrm{s}} \mathbb{E} \left\lbrace \left\Vert \tilde{\matr{s}}_t - \hat{\matr{s}}_t \right\Vert^2 \right\rbrace \mathrm{d}t$ in the following.  We compare the minimum achievable normalized time-averaged \gls{mse} of the proposed task-based \gls{adc} system to analog and digital recovery under a fixed rate budget $R = K \cdot f_\mathrm{s} \cdot b$. For the proposed system we perform a grid search over all feasible combinations of number of bits per sample $b \in \{1,\ldots,16\}$ and number of \glspl{adc} $K \in \{1,\ldots,N\}$. For the analog and digital recovery systems, the number of \glspl{adc} is fixed to $K=N$ and $K=M$, respectively. Hence, for those systems we only perform a grid search over $b$. The sampling rate is then chosen as $f_\mathrm{s} = \frac{R}{K \cdot b}$, because the time-averaged \gls{mse} is monotonically decreasing with increasing $f_\mathrm{s}$.\looseness-1

\begin{figure}
	\centering
	\includegraphics[width=\columnwidth]{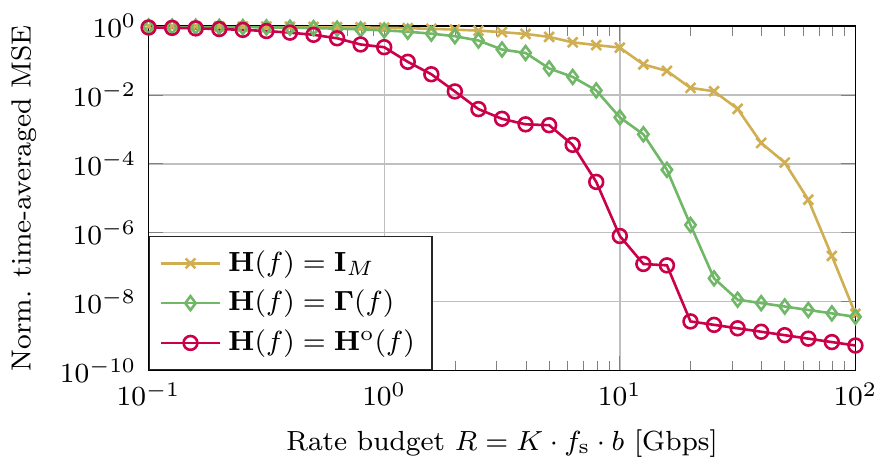}
	\caption{Achievable normalized time-averaged \gls{mse} for a fixed rate budget $R$. Solution is obtained by a grid search over $K \in \{1,\ldots,N\}$ and $b \in \{1, \ldots, 16\}$, where the sampling rate is chosen as $f_\mathrm{s} = \frac{R}{K \cdot b}$.\vspace{-0.4cm}}
	\label{fig:eval_mse_fixed_rate}
\end{figure}

In Fig.~\ref{fig:eval_mse_fixed_rate}, it can be observed that the proposed task-based \gls{adc} system outperforms the two competing architectures significantly, when operating under a constraint on the bit rate: To achieve a normalized \gls{mse} of $10^{-2}$, the proposed system requires only approx.\ \SI{25}{\percent} and approx.\ \SI{8}{\percent} of the rate as compared to analog and digital recovery systems, respectively. To achieve a normalized \gls{mse} of $10^{-4}$ the bit rate savings become smaller, i.e., the proposed system requires only approx.\ \SI{47}{\percent} and approx.\ \SI{14}{\percent} of the rate as compared to analog and digital recovery systems, respectively.

\begin{figure}
	\centering
	\includegraphics[width=0.95\columnwidth]{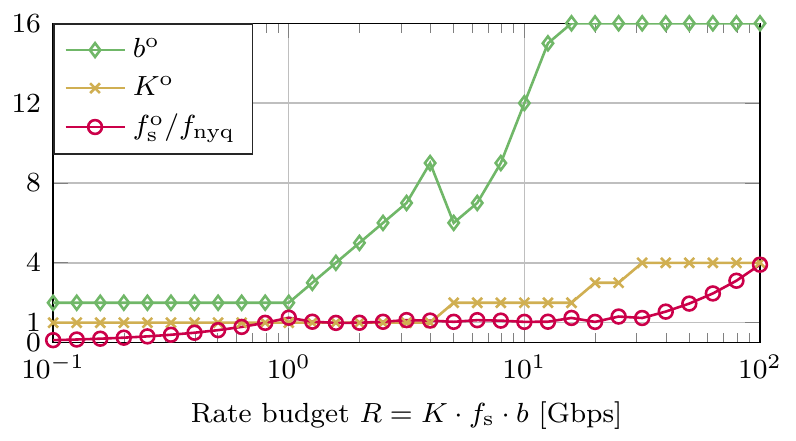}
	\caption{\gls{mse}-minimizing configuration of the proposed system which yields the performance depicted in Fig.~\ref{fig:eval_mse_fixed_rate}.
	\vspace{-0.3cm}}
	\label{fig:eval_opt_conf_journal}
\end{figure}

The \gls{mse}-minimizing   configuration $(K^\mathrm{o}, f_\mathrm{s}^\mathrm{o}, b^\mathrm{o})$ of the proposed system, which yields the performance depicted in Fig.~\ref{fig:eval_mse_fixed_rate}, is evaluated in Fig.~\ref{fig:eval_opt_conf_journal}. For very low rate budgets, \iec for $R < \SI{1}{Gbps}$, the \gls{mse}-minimizing system employs sub-Nyquist sampling, i.e., $f_\mathrm{s}^\mathrm{o} < f_\mathrm{nyq}$. This shows that under a strict rate budget employing sub-Nyquist sampling in order to allow for an increased amplitude resolution minimizes the \gls{mse}. Notably, $b^\mathrm{o} \geq 2$ holds for all considered rate budgets, \iec 1-bit quantization never minimizes the \gls{mse} for the considered task.
Moreover, in general we observe that with an increasing rate budget first the amplitude resolution $b$ and then the number of \glspl{adc} $K$ is increased.
For rate budgets $R \geq \SI{30}{Gbps}$, it holds that $K^\mathrm{o}=N$, \iec the number of \glspl{adc} corresponds to the dimension of the task, while for lower rates it is observed that  $K^\mathrm{o}< N$, in agreement with  Corollary~\ref{cor:optimal_K}.
Significant oversampling, i.e., $f_\mathrm{s}^\mathrm{o} \gg f_\mathrm{nyq}$ is only employed when $b$ cannot be increased anymore (due to the restriction $b \leq 16$) and $K$ is sufficient, i.e., $K = \max \rank{\bar{\mat{\Gamma}}(f)}$.

\vspace{-0.2cm}
\section{Conclusions}\label{sec:conclusions}
\vspace{-0.1cm}
In this work, we proposed hybrid analog-digital task-based \gls{adc} systems, where the analog and digital filters were optimized to minimize the \gls{mse} in recovering a linear task vector from a set of \gls{ct} zero-mean \gls{wss} input signals for a fixed sampling rate and quantizer resolution. We characterized the \gls{mse}-minimizing linear filters and obtained analytical expressions for the minimum achievable distortion. From our analytical results, we obtained design guidelines for practical systems. Moreover, our results prove that the proposed joint design is in general superior to implementing recovery solely in the analog or digital domain. In our numerical study, we utilized the analytical results to obtain the \gls{mse}-minimizing \gls{adc} configuration numerically under a constraint on the digital bit rate for the task of recovering the output of a \gls{mimo} matched filter. We showed that the proposed approach substantially outperforms analog and digital matched filtering. Additionally, we demonstrated numerically that sub-Nyquist sampling of bandlimited signals can improve the accuracy of recovering a task vector subject to a constraint on the overall bit rate.\looseness-10

\renewcommand{\theequation}{A\arabic{equation}}
\setcounter{equation}{0}  
\vspace{-0.2cm}
\begin{appendix}\label{sec:appendix}
\vspace{-0.2cm}
\subsection{Proof of Proposition \ref{th:optimal_digital_filter}}\label{sec:appendix_A}
\vspace{-0.1cm}
    To identify the \gls{mse}-minimizing digital filter, we use the orthogonality principle, which states that $\hat{\matr{s}} = \sum_{n\in \mathbb{Z}} \cev{\mat{G}}[n] \matr{z}[n]$ minimizes the \gls{mse} if and only if it holds that  \cite[eq.~(7-92)]{papoulis2001probability}
    \begin{equation}
        \mathbb{E} \Big\lbrace \left( \tilde{\matr{s}} - \hat{\matr{s}} \right) \big( \sum_{n\in \mathbb{Z}} \mat{A}[n]  \matr{z}[n] \big)^T \Big\rbrace = \mat{0}_{N}, \quad \forall \mat{A}[n] \in \mathbb{R}^{N \times K}.
        \label{eq:def_general_orth_principle}
    \end{equation}
    By writing $\mat{C}_{\matr{z}} [l] = \mathbb{E} \left\lbrace \matr{z}[n+l] \matr{z}^T[n] \right\rbrace$, this condition becomes
    \begin{equation}\label{eq:C_s_tilde_z_1}
          \sum_{n\in \mathbb{Z}} \mathbb{E} \left\lbrace \tilde{\matr{s}} \matr{z}^T[n]  \right\rbrace \mat{A}^T[n]  =  \sum_{n\in \mathbb{Z}} \sum_{l \in \mathbb{Z}} \cev{\mat{G}}[l] \cev{\mat{C}}_{\matr{z}}[n-l] \mat{A}^T[n],
    \end{equation}
 which in turn implies that 
    \begin{equation}
        \mathbb{E} \left\lbrace \tilde{\matr{s}} \matr{z}^T[n]  \right\rbrace = \sum_{l \in \mathbb{Z}} \cev{\mat{G}}[l] \cev{\mat{C}}_{\matr{z}}[n-l] = \left( \cev{\mat{G}} * \cev{\mat{C}}_{\matr{z}} \right) [n].
        \label{eq:def_C_s_tilde_z}
    \end{equation}
    Furthermore,  since  $ \matr{y}[n]  = T_\mathrm{s} \left( \mat{H * \matr{x}} \right)(nT_\mathrm{s})$, it follows
    \begin{align}
    	\label{eq:def_z_dep_x}
        \matr{z}[n] &=  T_\mathrm{s} \int_{\mathbb{R}} \mat{H}(\tau) \matr{x}(n T_\mathrm{s} - \tau) \mathrm{d}\tau + \matr{e}[n].
    \end{align}
    Inserting \eqref{eq:def_Gamma_t} and \eqref{eq:def_z_dep_x} into the \gls{lhs} of \eqref{eq:def_C_s_tilde_z} yields
    \begin{equation} 
        \mathbb{E} \left\lbrace \tilde{\matr{s}} \matr{z}^T[n]  \right\rbrace = \left( \cev{\mat{\Gamma}} * \cev{\mat{C}}_{\matr{x}} * \mat{H}^T \right) [n],
        \label{eq:C_s_tilde_z_2}
    \end{equation} 
    where
    $\mat{C}_{\matr{x}} (\tau) \triangleq  \mathbb{E} \left\lbrace \matr{x}(t+\tau) \matr{x}(t) \right\rbrace$.
    Combining \eqref{eq:def_C_s_tilde_z}-\eqref{eq:C_s_tilde_z_2} yields
    \begin{equation}
        \left( \cev{\mat{G}} * \cev{\mat{C}}_{\matr{z}} \right) [n] = \left( \cev{\mat{\Gamma}} * \cev{\mat{C}}_{\matr{x}} * \mat{H}^T \right) [n].
        \label{eq:convolution_equality}
    \end{equation}
    Taking the \gls{dtft} of \eqref{eq:convolution_equality} yields
    \begin{align}
        &\mat{G}^*(e^{j 2 \pi f T_\mathrm{s}}) \cdot \mat{C}_{\matr{z}}^*(e^{j 2 \pi f T_\mathrm{s}}) = \nonumber\\
        & \quad = \sum_{k \in \mathbb{Z}} \mat{\Gamma}^*\left(f - k f_\mathrm{s}\right) \mat{C}_{\matr{x}}^* \left(f - k f_\mathrm{s}\right) \mat{H}^T\left(f - k f_\mathrm{s}\right).
        \label{eq:help1}
    \end{align}
    Solving \eqref{eq:help1} for $\mat{G}$ and taking the inverse \gls{dtft} proves \eqref{eq:theorem_optimal_G_omega}.

    Next, we express the resulting \gls{mse}.
    For a given analog filter $\mat{H}(f)$, the \gls{mse} of the digital recovery filter $\mat{G}[n]$ is 
    \begin{equation} 
        \mse{\mat{H}(f)} =\E \left\lbrace \Vert \tilde{\matr{s}} \Vert^2  \right\rbrace - \trace{ \E \left\lbrace \tilde{\matr{s}} \hat{\matr{s}}^T  \right\rbrace }.
        \label{eq:def_mmse_wrt_s_tilde}
    \end{equation}
        Using  \eqref{eq:def_s_hat} and \eqref{eq:def_C_s_tilde_z}, it follows that
    \begin{align}
        &\E \left\lbrace \tilde{\matr{s}} \hat{\matr{s}}^T \right\rbrace  =    \left( \mat{G} * \mat{C}_{\matr{z}} * \cev{\mat{G}}^T \right) [0] \nonumber \\
        & =  T_\mathrm{s} \int_{-\frac{f_\mathrm{s}}{2}}^{\frac{f_\mathrm{s}}{2}} \mat{G}(e^{j 2 \pi f T_\mathrm{s}}) \, \mat{C}_{\matr{z}}(e^{j 2 \pi f T_\mathrm{s}})   \mat{G}^H(e^{j 2 \pi f T_\mathrm{s}}) \, \mathrm{d} f.
        \label{eq:E_s_tilde_s_hat}
    \end{align}
    Substituting $\mat{G}(e^{j 2 \pi f T_\mathrm{s}})$  from \eqref{eq:help1} into \eqref{eq:E_s_tilde_s_hat}, and inserting the result into \eqref{eq:def_mmse_wrt_s_tilde} yields \eqref{eq:theorem_mse_h_omega_general}, which concludes the proof.
\IEEEQED
\vspace{-0.2cm}
\subsection{Proof of Theorem \ref{th:optimal_pre_filter}}\label{sec:appendix_B}
\vspace{-0.1cm}
    We prove the theorem by finding the analog filter $\mat{H}(f)$ which minimizes the \gls{mse} expression from Proposition~\ref{th:optimal_digital_filter}, given by \eqref{eq:theorem_mse_h_omega_general}. Because $\E \left\lbrace \Vert \tilde{\matr{s}} \Vert^2  \right\rbrace$ is independent of $\mat{H}(f)$, the expression is minimized by maximizing the second term \wrt $\mat{H}(f)$.
    Therefore, we express the optimization solely \wrt $\mat{H}(f)$ first by formulating the relation between $\mat{H}(f)$ and the quantization error variance $\frac{\Delta^2}{4}$. Then, we use the resulting optimization problem to identify the matrices comprising \eqref{eq:theorem_H_opt}.\looseness-1

    Since $\tilde{\mathsf{y}}_k[n]$ is the sum of the independent variables $\mathsf{y}_k[n]$ and $\mathsf{w}_k[n]$, where  $\E \{ \mathsf{w}_k^2[n] \} = \frac{\Delta^2}{6}$ and $\Delta = \frac{2 \, \gamma}{2^b}$, it holds that \eqref{eq:def_dynamic_range_squared} can be written as
	$\gamma^2 =  \eta^2 \left( \max_{k \in \mathcal{K}} \mathbb{E} \left\lbrace \mathsf{y}^2_k[n] \right\rbrace + \frac{2 \, \gamma^2}{3 \, 2^{2b}} \right)$, thus
    \begin{equation} 
          \gamma^2 = \bar{\kappa} \, \max_{k \in \mathcal{K}} \, \mathbb{E}\{ \mathsf{y}_k^2[n] \} =  \bar{\kappa} \, \max_{k \in \mathcal{K}} \left[ \mat{C}_{\matr{y}}[0] \right]_{k,k},
        \label{eq:mod_def_dynamic_range_squared}
    \end{equation}
    with $\bar{\kappa} = \eta^2 (1-\frac{2 \, \eta^2}{3 \, 2^{2b}})^{-1}$, and 
    \begin{equation} 
        \mat{C}_{\matr{y}}[0] \! = \! T_\mathrm{s}^2 \int_{-\frac{f_\mathrm{s}}{2}}^{\frac{f_\mathrm{s}}{2}} \sum_{k \in \mathbb{Z}} \mat{H}_k \left( f \right) \mat{C}_{\matr{x},k} \left( f \right)   \mat{H}_k^H\left( f  \right) \mathrm{d}f,
        \label{eq:def_C_y_0}
    \end{equation}
    where we used the shorthand notations $\mat{H}_k(f) \triangleq \mat{H}(f - k f_\mathrm{s})$ and $\mat{C}_{\matr{x},k}(f) \triangleq \mat{C}_{\matr{x}}(f - k f_\mathrm{s})$.
    Inserting \eqref{eq:mod_def_dynamic_range_squared} and $\Delta = \frac{2 \, \gamma}{2^b}$ into \eqref{eq:def_C_e} and subsequently applying the \gls{dtft} yields
    \begin{equation}
        \mat{C}_{\matr{e}}(e^{j 2 \pi f T_\mathrm{s}}) = \kappa \, \max_{k \in \mathcal{K}} \left[ \mat{C}_{\matr{y}}[0] \right]_{k,k} \mat{I}_K,
        \label{eq:def_C_e_f}
    \end{equation}
    with $\kappa = \frac{\bar{\kappa}}{2^{2b}}$. Then, from \eqref{eq:def_z_dep_x} it follows that
    $\mat{C}_{\matr{z}}(e^{j 2 \pi f T_\mathrm{s}}) \!= \! T_\mathrm{s} \sum_{k \in \mathbb{Z}} \mat{H}_k \!\left(  f \right) \mat{C}_{\matr{x},k} \!\left(  f  \right) \mat{H}_k^H\!\left(  f  \right)\! +\! \mat{C}_{\matr{e}}(e^{j 2 \pi f T_\mathrm{s}})$. 
 	Combining this expression for $\mat{C}_{\matr{z}}(\cdot)$ with \eqref{eq:def_C_y_0}, \eqref{eq:def_C_e_f} and \eqref{eq:theorem_mse_h_omega_general}, 
    as well as the definitions definitions \eqref{eq:def_Gamma_bar}-\eqref{eq:def_H_bar} and the fact that  $\sum_{k = - \Upsilon}^{\Upsilon} \mat{\Gamma}_k(f)  \mat{C}_{\matr{x},k}(f) \mat{H}^H_k(f) = \bar{\mat{\Gamma}}(f) \bar{\mat{H}}^H(f)$,   we formulate the optimization of the analog filter $\mat{H}(f)$ as the maximization problem given in \eqref{eq:max_trace_term_H_bar}.  
	\begin{figure*}[ht!]
	\begin{IEEEeqnarray}{Cll}
		\max_{\bar{\mat{H}}(f)} & \mathrm{tr} \bigg(  \int_{-\frac{f_\mathrm{s}}{2}}^{\frac{f_\mathrm{s}}{2}} & \bar{\mat{\Gamma}}(f) \bar{\mat{H}}^H(f) \left( \bar{\mat{H}}(f) \bar{\mat{H}}^H(f) + \kappa T_\mathrm{s} \max_{k \in \mathcal{K}} \left[ \int_{-\frac{f_\mathrm{s}}{2}}^{\frac{f_\mathrm{s}}{2}} \bar{\mat{H}}(f') \bar{\mat{H}}^H(f') \mathrm{d}f' \right]_{k,k} \mat{I}_K \right)^{-1} \bar{\mat{H}}(f) \bar{\mat{\Gamma}}^H(f) \mathrm{d} f \bigg) \qquad
		\label{eq:max_trace_term_H_bar}
		\\
		\! \max_{\bar{\mat{\Sigma}}(f),\bar{\mat{V}}(f)} & \!\mathrm{tr} \!\bigg( \! \int_{-\frac{f_\mathrm{s}}{2}}^{\frac{f_\mathrm{s}}{2}}\! & \bar{\mat{\Gamma}}(f) \bar{\mat{V}}(f) \bar{\mat{\Sigma}}^H\!(f)\! \left( \!\bar{\mat{\Sigma}}(f) \bar{\mat{\Sigma}}^H\!(f) \!+\! \frac{\kappa T_\mathrm{s}}{K} \trace{\! \int_{-\frac{f_\mathrm{s}}{2}}^{\frac{f_\mathrm{s}}{2}}\! \bar{\mat{\Sigma}}(f') \bar{\mat{\Sigma}}^H(f') \mathrm{d}f'\!} \mat{I}_K\! \right)^{-1}\!\! \bar{\mat{\Sigma}}(f) \bar{\mat{V}}^H\!(f) \bar{\mat{\Gamma}}^H\!(f) \mathrm{d} f \bigg) \qquad
		\label{eq:max_trace_term_wrt_Lambda_V}
	\end{IEEEeqnarray}
	\hrulefill
	\vspace*{4pt}
	\vspace{-0.2cm}
	\end{figure*}
   Next, we use the \gls{svd}:
   \vspace{-0.1cm}
    \begin{equation}
        \bar{\mat{H}}(f) = \bar{\mat{U}}(f) \bar{\mat{\Sigma}}(f) \bar{\mat{V}}^H(f),
        \label{eq:svd_H_bar}
   \vspace{-0.1cm}
    \end{equation}
    where $\bar{\mat{U}}(f) \in \mathbb{C}^{K \times K}$ and $\bar{\mat{V}}(f) \in \mathbb{C}^{\bar{M} \times \bar{M}}$ are unitary matrices and $\bar{\mat{\Sigma}}(f) \in \mathbb{R}^{K \times \bar{M}}$ is a diagonal matrix containing the singular values of $\bar{\mat{H}}(f)$ in descending order on its diagonal.
    When inserting \eqref{eq:svd_H_bar} into \eqref{eq:max_trace_term_H_bar}, it can be seen that $\bar{\mat{U}}(f)$ only affects the term
    \begin{equation*}
        \rho \triangleq \kappa T_\mathrm{s} \max_{k \in \mathcal{K}} \big[ \int_{-\frac{f_\mathrm{s}}{2}}^{\frac{f_\mathrm{s}}{2}} \bar{\mat{U}}(f') \bar{\mat{\Sigma}}(f') \bar{\mat{\Sigma}}^H(f') \bar{\mat{U}}^H(f') \mathrm{d}f' \big]_{k,k}.
    \end{equation*}
    Then, following similar steps as in \cite[Appendix~C]{shlezinger2018hardwarelimited}, it holds that for any pair of positive semi-definite matrices $\mat{M}_1$, $\mat{M}_2$, the scalar function $f(\rho) = \trace{\mat{M}_1(\mat{M}_2 + \rho \mat{I})^{-1}}$ is monotonically decreasing with $\rho$ for $\rho > 0$. Hence, $f(\rho)$ is maximized by minimizing $\rho$. Using \cite[Corollary 2.4]{palomar2007mimo}, it follows that
    \vspace{-0.1cm}
    \begin{IEEEeqnarray}{ll}
        \min_{\bar{\mat{U}}(f)} \, & \max_{k \in \mathcal{K}} \left[ \int_{-\frac{f_\mathrm{s}}{2}}^{\frac{f_\mathrm{s}}{2}} \bar{\mat{U}}(f') \bar{\mat{\Sigma}}(f') \bar{\mat{\Sigma}}^H(f') \bar{\mat{U}}^H(f') \mathrm{d}f' \right]_{k,k} \nonumber\\
        & = \frac{1}{K} \trace{ \int_{-\frac{f_\mathrm{s}}{2}}^{\frac{f_\mathrm{s}}{2}} \bar{\mat{\Sigma}}(f') \bar{\mat{\Sigma}}^H(f') \mathrm{d}f'},
        \label{eq:min_trace}
    \vspace{-0.1cm}
    \end{IEEEeqnarray}
    where the minimizing unitary matrix $\bar{\mat{U}}$, denoted as $\bar{\mat{U}}_{\mat{H}}(f)$, can be obtained using \cite[Algorithm 2.2]{palomar2007mimo}.
    
    Substituting \eqref{eq:min_trace} 
    and \eqref{eq:svd_H_bar} into \eqref{eq:max_trace_term_H_bar}
    yields \eqref{eq:max_trace_term_wrt_Lambda_V}. Noting that \eqref{eq:max_trace_term_wrt_Lambda_V} is invariant to any scalar scaling of $\bar{\mat{\Sigma}}(f)$, i.e., replacing $\bar{\mat{\Sigma}}(f)$ with $c\bar{\mat{\Sigma}}(f)$, $\forall c \in \mathbb{R}$, we can rewrite the unconstrained optimization problem \eqref{eq:max_trace_term_wrt_Lambda_V} as an equivalent constrained optimization problem:
    \vspace{-0.1cm}
    \begin{subequations}
        \label{eq:max_trace_term_equiv_constrained}
        \begin{IEEEeqnarray}{ll}
            \label{eq:max_trace_term_equiv_constraind_objective}
            \max_{\bar{\mat{\Sigma}}(f),\bar{\mat{V}}(f)}  & \mathrm{tr}\! \left( \! \int_{-\frac{f_\mathrm{s}}{2}}^{\frac{f_\mathrm{s}}{2}\!} \bar{\mat{\Gamma}}(f) \bar{\mat{V}}(f) \bar{\mat{\Sigma}}^H\!(f)\!
            \left( \!\bar{\mat{\Sigma}}(f) \bar{\mat{\Sigma}}^H\!(f)\! +\! \frac{1}{2^{2b}}\mat{I}_K \right)^{-1} \right. \nonumber \\
            & \left. \vphantom{\int_{-\frac{f_\mathrm{s}}{2}}^{\frac{f_\mathrm{s}}{2}}} \quad \times \bar{\mat{\Sigma}}(f) \bar{\mat{V}}^H(f) \bar{\mat{\Gamma}}^H(f) \mathrm{d} f \right)\\ \label{eq:max_trace_term_equiv_constraind_equality_const}
            \mathrm{s.t.} & \frac{\bar{\kappa} T_\mathrm{s}}{K} \trace{ \int_{-\frac{f_\mathrm{s}}{2}}^{\frac{f_\mathrm{s}}{2}} \bar{\mat{\Sigma}}(f') \bar{\mat{\Sigma}}^H(f') \mathrm{d}f'} = 1. 
        \end{IEEEeqnarray}
    \vspace{-0.1cm}
    \end{subequations}
    
    Note that the equality constraint \eqref{eq:max_trace_term_equiv_constraind_equality_const} is invariant to $\bar{\mat{V}}(f)$. Hence, maximizing \eqref{eq:max_trace_term_equiv_constrained} \wrt $\bar{\mat{V}}(f)$ yields the following unconstrained maximization problem
   \vspace{-0.1cm}
    \begin{IEEEeqnarray}{lCl}
            \bar{\mat{V}}_{\mat{H}}(f) & = & \argmax_{\bar{\mat{V}}(f)} \, \int_{-\frac{f_\mathrm{s}}{2}}^{\frac{f_\mathrm{s}}{2}} \mathrm{tr} \left( \mat{B}(f) \mat{D}(f) \right) \mathrm{d} f,
            \label{eq:argmax_V}
   \vspace{-0.1cm}
    \end{IEEEeqnarray}
    with $\mat{B}(f)  \triangleq \bar{\mat{V}}^H(f) \bar{\mat{\Gamma}}^H\left( f \right) \bar{\mat{\Gamma}}\left(f \right) \bar{\mat{V}}(f)$ and $\mat{D}(f)  \triangleq  \bar{\mat{\Sigma}}^H(f) \left( \bar{\mat{\Sigma}}(f) \bar{\mat{\Sigma}}^H(f) + \frac{1}{2^{2b}}\mat{I}_K \right)^{-1} \bar{\mat{\Sigma}}(f)$. The matrix  
    $\mat{D}(f) $ is diagonal with descending  entries  $\left[ \mat{D}(f) \right]_{i,i} = \frac{ \left[ \bar{\mat{\Sigma}}(f) \right]_{i,i}^2 }{ \left[ \bar{\mat{\Sigma}}(f) \right]_{i,i}^2 + \frac{1}{2^{2b}} }, i \in \bar{\mathcal{M}}\triangleq\{1, \ldots, \bar{M}\}$. From \cite[Th.~II.1]{lasserre1995A_trace} it follows that
    \vspace{-0.1cm}
    \begin{equation}
        \label{eq:trace_eigenvalue_inequality}
        \mathrm{tr} \left( \mat{B}(f) \mat{D}(f) \right) \leq \sum_{i=1}^{\bar{M}} \lambda_{\mat{B}(f),i} \, \lambda_{\mat{D}(f),i},
    \vspace{-0.1cm}
    \end{equation}
    where $\lambda_{\mat{B}(f),i}$ and $\lambda_{\mat{D}(f),i} = \left[ \mat{D}(f) \right]_{i,i}$, $i \in \bar{\mathcal{M}}$, denote the $i$th eigenvalue of $\mat{B}(f)$ and $\mat{D}(f)$, respectively.
    From the \emph{rearrangement inequality} \cite[Th.~368]{hardy1934inequalities}, it follows that the \gls{rhs} of \eqref{eq:trace_eigenvalue_inequality} is maximized if the eigenvalues $\lambda_{\mat{B}(f),i}$ are also in descending order. In this case, the upper-bound in \eqref{eq:trace_eigenvalue_inequality} is tight.
    Hence, $\bar{\mat{V}}_{\mat{H}}(f)$ is the matrix of right-singular vectors of $\bar{\mat{\Gamma}}(f)$, \iec by writing the \gls{svd} of $\bar{\mat{\Gamma}}(f)$ as $ \bar{\mat{\Gamma}}(f) = \bar{\mat{U}}_\mat{\Gamma}(f) \bar{\mat{\Sigma}}_\mat{\Gamma}(f) \bar{\mat{V}}_\mat{\Gamma}^H(f)$ in which the diagonal entries of $\bar{\mat{\Sigma}}_\mat{\Gamma}(f)$ are sorted in descending order, it holds that $\bar{\mat{V}}_{\mat{H}}(f) = \bar{\mat{V}}_{\mat{\Gamma}}(f)$.
   Substituting $\bar{\mat{V}}(f) = \bar{\mat{V}}_{\mat{H}}(f)$ into \eqref{eq:max_trace_term_equiv_constrained} and noting that all resulting matrices are diagonal, we obtain
    \vspace{-0.1cm}
    \begin{subequations}
        \label{eq:equiv_constraint_op_max_scalar}%
        \begin{IEEEeqnarray}{ll}
            \label{eq:equiv_constrained_op_max_scalar_objective}%
            \max_{\alpha_i(f) \geq 0} \quad & \int_{-\frac{f_\mathrm{s}}{2}}^{\frac{f_\mathrm{s}}{2}} \sum_{i=1}^{\min(K,\bar{M})} \frac{\sigma_{\bar{\mat{\Gamma}},i}^2(f) \alpha_i(f)}{\alpha_i(f) + \frac{1}{2^{2b}}} \, \mathrm{d} f \hphantom{aaaaaaaa}\\
            \label{eq:equiv_constrained_op_max_scalar_constraint}%
            \mathrm{s.t.} & \frac{\bar{\kappa} \, T_\mathrm{s}}{K} \, \int_{-\frac{f_\mathrm{s}}{2}}^{\frac{f_\mathrm{s}}{2}} \sum_{i=1}^{\min(K,\bar{M})} \alpha_i(f) \, \mathrm{d} f = 1,
        \end{IEEEeqnarray}%
    \vspace{-0.1cm}%
    \end{subequations}
    where $\sigma_{\bar{\mat{\Gamma}},i}^2(f) = \left[ \bar{\mat{\Sigma}}_{\mat{\Gamma}}(f) \right]_{i,i}^2$ and $\alpha_i(f) = \left[ \bar{\mat{\Sigma}}(f) \right]_{i,i}^2$, $i \in \bar{\mathcal{M}}$.
    The function $\frac{\sigma_{\bar{\mat{\Gamma}},i}^2(f) \alpha_i(f)}{\alpha_i(f) + \frac{1}{2^{2b}}}$ is concave in $\alpha_i(f)$ for $\alpha_i(f) \geq 0$.
    Hence, the objective \eqref{eq:equiv_constrained_op_max_scalar_objective} is concave in $\alpha_i(f)$, as the sum and integral operations preserve concavity \cite[Sec.~3.2.1]{boyd2004convex}. 
    Furthermore, since the objective and constraint of \eqref{eq:equiv_constraint_op_max_scalar} are differentiable, the \gls{kkt} conditions are necessary and sufficient for optimality \cite[Sec.~5.5.3]{boyd2004convex}.
    Solving the \gls{kkt} conditions of \eqref{eq:equiv_constraint_op_max_scalar} yields the solution
    \vspace{-0.1cm}
    \begin{IEEEeqnarray}{lCl}
        \alpha_i(f) & = & \left( \sqrt{\frac{K}{2^{2b} \, \bar{\kappa} \, T_\mathrm{s} \, \mu}} \sigma_{\bar{\mat{\Gamma}},i}(f) - \frac{1}{2^{2b}} \right)^+,
        \label{eq:alpha_i_opt}
    \vspace{-0.1cm}
    \end{IEEEeqnarray}
    where $\mu$ is set such that \eqref{eq:equiv_constrained_op_max_scalar_constraint} holds. As $\mu$ scales $\sigma_{\bar{\mat{\Gamma}},i}(f)$ in \eqref{eq:alpha_i_opt}, we can simplify \eqref{eq:alpha_i_opt} to  
    \begin{equation}
        \label{eq:alpha_i_opt_simple}
        \alpha_i(f) = \frac{1}{2^{2b}} \left( \zeta \sigma_{\bar{\mat{\Gamma}},i}(f) - 1 \right)^+, 
    \end{equation}
    where now $\zeta$ has to be chosen such that  \eqref{eq:equiv_constrained_op_max_scalar_constraint} holds. Using \eqref{eq:alpha_i_opt_simple}, the \gls{mse}-minimizing matrix of singular values, denoted as $\bar{\mat{\Sigma}}_{\mat{H}}(f)$, is 
   \vspace{-0.1cm}
    \begin{equation}
        \label{eq:Lambda_H_opt}
        \left[\bar{\mat{\Sigma}}_{\mat{H}}(f)\right]_{i,i} = \sqrt{\alpha_i(f)},  \quad i \in \{1, \ldots, \min(K,\bar{M})\}.
   \vspace{-0.1cm}
    \end{equation}
    Finally, inserting  $\bar{\mat{V}}(f) = \bar{\mat{V}}_{\mat{H}}(f)$ and \eqref{eq:Lambda_H_opt} into \eqref{eq:svd_H_bar} and utilizing \eqref{eq:min_trace} proves \eqref{eq:theorem_H_opt}.
    Furthermore, the objective  
    \eqref{eq:equiv_constrained_op_max_scalar_objective} equals the trace  in the \gls{rhs} of \eqref{eq:theorem_mse_h_omega_general}. Hence, inserting \eqref{eq:Lambda_H_opt} into  
    \eqref{eq:equiv_constrained_op_max_scalar_objective} and inserting the result into the \gls{rhs} of \eqref{eq:theorem_mse_h_omega_general} yields \eqref{eq:theorem_mse_H_opt}, hence, concluding the proof.
\IEEEQED

\vspace{-0.2cm}
\subsection{Proof of Corollary \ref{cor:optimal_K} and \ref{cor:trade-off_fs_b}}\label{sec:appendix_C}
\vspace{-0.1cm}
    First, noting that the \gls{mse} expression given by \eqref{eq:theorem_mse_H_opt} is monotonically decreasing in $\zeta$, we conclude that $\zeta$ should be maximized.
    
    Then, we begin by proving Corollary \ref{cor:optimal_K}. Let $L$ denote the maximum number of non-zero singular values $\sigma_{\bar{\mat{\Gamma}},i}(f)$ of $\bar{\mat{\Gamma}}(f)$. By definition, $L$ is the maximal value of $\rank{\bar{\mat{\Gamma}}(f)}$ over $f \in \big[ -\frac{f_\mathrm{s}}{2}, \frac{f_\mathrm{s}}{2} \big]$.  For $K \geq L$, it holds that $\sigma_{\bar{\mat{\Gamma}},i}(f)=0$ for $i > L$, and thus, by \eqref{eq:theorem_1_eq_const}, $\zeta$ is chosen such that
    \begin{align}
        & \frac{\bar{\kappa} \, T_\mathrm{s} }{K \, 2^{2b}} \int_{-\frac{f_\mathrm{s}}{2}}^{\frac{f_\mathrm{s}}{2}} \sum_{i=1}^{\min(L,\bar{M})} \left( \zeta \sigma_{\bar{\mat{\Gamma}},i}(f) - 1 \right)^+ \, \mathrm{d} f = 1.
        \label{eq:corollary_2a}
    \end{align} 
    Using $\bar{\kappa} = \eta^2 (1-\frac{2 \, \eta^2}{3 \, 2^{2b}})^{-1}$, \eqref{eq:corollary_2a} implies that 
    \begin{equation}
    \label{eq:corollary_2}
    \! \! \int_{\!-\!\frac{f_\mathrm{s}}{2}}^{\frac{f_\mathrm{s}}{2}} \!\sum_{i=1}^{\min(L,\bar{M})} \!\!\!\left( \zeta \sigma_{\bar{\mat{\Gamma}},i}(f) \!-\! 1 \right)^+ \, \!\mathrm{d} f \!= \! K f_{\rm s}\left(\frac{2^{2b}}{\eta^2 } - \frac{2}{3} \right).
    \end{equation}
    The \gls{rhs} of \eqref{eq:corollary_2} should be maximized in order to maximize $\zeta$. Furthermore, under a fixed rate budget $R=K \cdot f_\mathrm{s} \cdot b$, we can define the quantizer resolution as 
    $b = \left\lfloor \frac{R}{K \cdot f_\mathrm{s}} \right\rfloor$,
    where we restrict ourselves to combinations of $K$ and $f_\mathrm{s}$, such that $b \geq 1$. Now substituting this expression for $b$ into the \gls{rhs} of \eqref{eq:corollary_2} yields an expression which is monotonically decreasing in $K$ for a given $f_{\rm s}$. As a result, for $K \geq L$, the \gls{mse} expression \eqref{eq:theorem_mse_H_opt} is minimized for $K = L$, which concludes the proof of Corollary \ref{cor:optimal_K}.\looseness-10
    
    Next we prove Corollary \ref{cor:trade-off_fs_b}. Again we insert $b = \left\lfloor \frac{R}{K \cdot f_\mathrm{s}} \right\rfloor \geq 1$ into the \gls{rhs} of \eqref{eq:corollary_2} in order to maximize $\zeta$. However, now for fixed $K$. Thereby, we obtain an expression which is monotonically decreasing in $f_\mathrm{s}$. Hence, $\zeta$ is maximized by maximizing $b$, which concludes the proof of Corollary \ref{cor:trade-off_fs_b}.
\IEEEQED

\vspace{-0.2cm}
\subsection{Proof of Corollary \ref{cor:analog_recovery_opt}}\label{sec:appendix_D}
\vspace{-0.1cm}
We seek necessary and sufficient conditions for $\mat{H}^\mathrm{o}(f) = \mat{\Gamma}(f)$, which implies $K=N$.
For $\mat{H}^\mathrm{o}(f) = \mat{\Gamma}(f)$ it follows from \eqref{eq:def_Gamma_bar} and \eqref{eq:def_H_bar} that
$\bar{\mat{\Gamma}}(f) = \bar{\mat{H}}(f) =\bar{\mat{U}}_{\mat{H}}(f) \bar{\mat{\Sigma}}_{\mat{H}}(f) \bar{\mat{V}}^H_{\mat{H}}(f)$.
Now, $\bar{\mat{V}}_{\mat{H}}(f)$ already equals the matrix of right-singular vectors of $\bar{\mat{\Gamma}}(f)$. Therefore, $\mat{H}(f)=\mat{\Gamma}(f)$ is optimal if and only if $\bar{\mat{\Gamma}}(f) \bar{\mat{\Gamma}}^H(f) = \bar{\mat{U}}_{\mat{H}}(f) \bar{\mat{\Sigma}}_{\mat{H}}(f) \bar{\mat{\Sigma}}_{\mat{H}}^H(f) \bar{\mat{U}}_{\mat{H}}^H(f)$. This implies that 
$ \left[ \bar{\mat{\Sigma}}_{\mat{H}}(f) \right]_{i,i}^2 = \sigma_{\bar{\mat{\Gamma}},i}^2(f)$, for each $i \in \mathcal{K}$ and $|f|<\frac{f_{\mathrm{s}}}{2}$,
where $\sigma_{\bar{\mat{\Gamma}},i}(f)$ denotes the $i$th singular value of $\bar{\mat{\Gamma}}(f)$. If $\bar{\mat{\Gamma}}(f) \bar{\mat{\Gamma}}^H(f)$ is non-singular for all $|f|<\frac{f_{\mathrm{s}}}{2}$, it follows $\sigma_{\bar{\mat{\Gamma}},i}(f) \neq 0$, $\forall i\in \mathcal{K}$, $\forall |f|<\frac{f_{\mathrm{s}}}{2}$. Thus, using \eqref{eq:theorem_1_def_Sigma} yields
\begin{equation}
    \label{eq:sigma_Gamma_bar_opt}
    \sigma_{\bar{\mat{\Gamma}},i}^2(f) = \frac{1}{2^{2b}} \left( \zeta \cdot \sigma_{\bar{\mat{\Gamma}},i}(f) - 1 \right)^+ = \frac{\zeta \cdot \sigma_{\bar{\mat{\Gamma}},i}(f) - 1}{2^{2b}} .
\end{equation}
Hence, $\zeta \cdot \sigma_{\bar{\mat{\Gamma}},i}(f) > 1$ for all $i \in \mathcal{K}$. Moreover, due to \eqref{eq:sigma_Gamma_bar_opt}: 
$  \sigma_{\bar{\mat{\Gamma}},i}(f) = \frac{1}{2^{2b+1}} \left( \zeta\pm \sqrt{\zeta^2 - 4^{b+1}} \right) \equiv \sigma_{\bar{\mat{\Gamma}}}$,
\iec the singular values $\sigma_{\bar{\mat{\Gamma}},i}(f)$ are identical for all $i \in \mathcal{K}$ and for all \hbox{$f \in \big[ -\frac{f_\mathrm{s}}{2}, \frac{f_\mathrm{s}}{2}\big]$}.
Hence, $\bar{\mat{\Sigma}}_{\mat{H}}(f) \bar{\mat{\Sigma}}_{\mat{H}}^H(f) = \sigma^2_{\bar{\mat{\Gamma}}} \mat{I}_K$. Moreover, $\bar{\mat{U}}_{\mat{H}}(f) = \bar{\mat{U}}_{\mat{H}}$ can be an arbitrary unitary matrix and both conditions on $\bar{\mat{U}}_{\mat{H}}(f)$, i.e., the condition in Theorem~\ref{th:optimal_pre_filter} and the condition $\bar{\mat{\Gamma}}(f) \bar{\mat{\Gamma}}^H(f) = \bar{\mat{U}}_{\mat{H}}(f) \bar{\mat{\Sigma}}_{\mat{H}}(f) \bar{\mat{\Sigma}}_{\mat{H}}^H(f) \bar{\mat{U}}_{\mat{H}}^H(f)$, are satisfied. As a result, analog recovery is optimal if and only if 
$\bar{\mat{\Gamma}}(f) \bar{\mat{\Gamma}}^H(f) = \sigma^2_{\bar{\mat{\Gamma}}} \, \mat{I}_N$. 
Furthermore, from \eqref{eq:def_Gamma_bar} it follows
\vspace{-0.1cm}
\begin{equation}
    \bar{\mat{\Gamma}}(f) \bar{\mat{\Gamma}}^H(f) = \sum_{k=-\Upsilon}^{\Upsilon} \mat{\Gamma}_k(f) \mat{C}_{\matr{x},k}(f) \mat{\Gamma}_k^H(f) = \sigma^2_{\bar{\mat{\Gamma}}} \, \mat{I}_N,
    \label{eq:Gamma_bar_outer_product_as_sum}
\vspace{-0.1cm}
\end{equation}
where $\mat{\Gamma}_k(f) = \mat{\Gamma}(f - k f_\mathrm{s})$.
Moreover, the covariance matrix of the analog \gls{mmse} estimate $\tilde{\matr{s}}$ is given by (cf. \eqref{eq:def_Gamma_t})
\begin{equation*}
    \mat{C}_{\tilde{\matr{s}}} = \left( \mat{\Gamma} * \mat{C}_{\matr{x}} * \cev{\mat{\Gamma}}^T \right)(0)  \overset{\mathrm{(a)}}{=}  \int_{-\frac{f_\mathrm{s}}{2}}^{{\frac{f_\mathrm{s}}{2}}} \sigma^2_{\bar{\mat{\Gamma}}} \, \mat{I}_N \, \mathrm{d}f = \underbrace{f_\mathrm{s} \sigma^2_{\bar{\mat{\Gamma}}} }_{= c} \, \mat{I}_N,
\end{equation*}
where (a) is due to \eqref{eq:Gamma_bar_outer_product_as_sum}. This concludes the proof.
\IEEEQED

\vspace{-0.2cm}
\subsection{Proof of Lemmas  \ref{lem:asym_MSE_Ts}-\ref{lem:asym_MSE_b}}\label{sec:appendix_E}
\vspace{-0.1cm}
In our proof of the individual lemmas we use the fact that the variance of the \gls{mmse} estimate $\tilde{\matr{s}}$ can be written as
\begin{equation} 
    \E \left\lbrace \Vert \tilde{\matr{s}} \Vert^2 \right\rbrace 
   = \trace{ \int_{-\frac{-f_\mathrm{s}}{2}}^{\frac{f_\mathrm{s}}{2}}  \bar{\mat{\Gamma}}(f) \bar{\mat{\Gamma}}^H(f) \mathrm{d}f }. 
    \label{eqn:exp_squared_norm_sTilde}
\end{equation}
We first prove Lemma \ref{lem:asym_MSE_Ts}. To that aim, we note that using \eqref{eq:def_Gamma_bar}-\eqref{eq:def_H_bar} combined with \eqref{eq:def_C_y_0}-\eqref{eq:def_C_e_f}, the \gls{mse}  \eqref{eq:theorem_mse_h_omega_general} becomes
\begin{align}
   & \mse{\mat{H}(f)} \!=\! \E \left\lbrace \Vert \tilde{\matr{s}} \Vert^2  \right\rbrace \!-\!   \mathrm{tr} \bigg( \! \int_{-\frac{f_\mathrm{s}}{2}}^{\frac{f_\mathrm{s}}{2}} \!\bar{\mat{\Gamma}}(f) \bar{\mat{H}}^H\!(f)\Big( \bar{\mat{H}}(f) \bar{\mat{H}}^H\!(f)   \nonumber \\
    & \qquad  + \kappa T_\mathrm{s} \max_{k \in \mathcal{K}} \left[ \tilde{\mat{C}}_{\matr{y}}[0] \right]_{k,k} \mat{I}_K \Big)^{-1}  \bar{\mat{H}}(f) \bar{\mat{\Gamma}}^H(f) \mathrm{d} f \bigg), 
    \label{eqn:MSEexpr}
\end{align}
where $\kappa = \frac{\bar{\kappa}}{2^{2b}}$ and $\tilde{\mat{C}}_{\matr{y}}[l] = \frac{1}{T_\mathrm{s}^2} \mat{C}_{\matr{y}}[l]$.
Recall that we focus on bandlimited inputs, as we did in our derivation of Theorem~\ref{th:optimal_pre_filter}. Consequently, for $T_\mathrm{s} \rightarrow 0$, the Shannon-Nyquist sampling theorem is satisfied for $T_\mathrm{s}$ small enough. 
In this case, neither $\bar{\mat{\Gamma}}(f)$ nor $\bar{\mat{H}}(f)$ depend on $T_\mathrm{s}$ (cf.~Corollary~\ref{cor:H_opt_Nyquist}), and thus taking  $T_\mathrm{s} \rightarrow 0$ in \eqref{eqn:MSEexpr} and using \eqref{eq:condition_H_maintains_sufficiency} results in
$\mse{\mat{H}(f)}  \xrightarrow{T_\mathrm{s} \rightarrow 0}    \E \left\lbrace \Vert \tilde{\matr{s}} \Vert^2  \right\rbrace - \mathrm{tr} \big(  \int_{-\frac{f_\mathrm{s}}{2}}^{\frac{f_\mathrm{s}}{2}} \bar{\mat{\Gamma}}(f) \bar{\mat{\Gamma}}^H(f) \mathrm{d} f \big)$,
which tends to zero by \eqref{eqn:exp_squared_norm_sTilde}.
This proves Lemma \ref{lem:asym_MSE_Ts}.

Next, we prove Lemma \ref{lem:asym_MSE_b}. Here, we note that
$ \kappa =   \frac{\eta^2}{2^{2b}} \left(1-\frac{2 \, \eta^2}{3 \, 2^{2b}}\right)^{-1}  \xrightarrow{b \rightarrow \infty} 0$.  Consequently, letting $b \rightarrow \infty$ in \eqref{eqn:MSEexpr} and using  the condition \eqref{eq:condition_H_maintains_sufficiency}, we obtain 
$  \mse{\mat{H}(f)}   \xrightarrow{b \rightarrow \infty}    \E \left\lbrace \Vert \tilde{\matr{s}} \Vert^2  \right\rbrace - \mathrm{tr} \big(  \int_{-\frac{f_\mathrm{s}}{2}}^{\frac{f_\mathrm{s}}{2}} \bar{\mat{\Gamma}}(f) \bar{\mat{\Gamma}}^H(f) \mathrm{d} f \big)$, which  tends to zero by \eqref{eqn:exp_squared_norm_sTilde}, thus, concluding the proof.
Note that this result remains valid when employing $\eta(b)=0.25b+1.75$ instead of a fixed $\eta$.
\IEEEQED

\end{appendix}

\bibliographystyle{IEEEtran}
\bibliography{ref_clean}

\begin{IEEEbiography}
    [{\includegraphics[width=1in,clip,keepaspectratio]{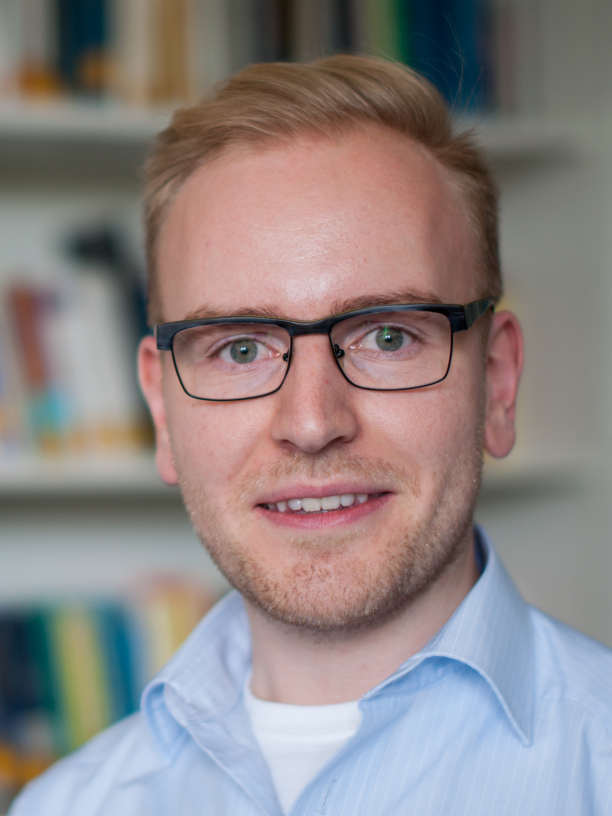}}]{Peter Neuhaus} (Graduate Student Member, IEEE) received the B.Sc. degree and the M.Sc. degree (with distinction) in electrical engineering, information technology, and computer engineering from RWTH Aachen University, Germany, in 2014 and 2017, respectively. He is currently pursuing the Ph.D. degree at the Vodafone Chair Mobile Communications Systems at Technische Universität Dresden, Germany. His research interests lie in the areas of signal processing and communications.
\end{IEEEbiography}

\begin{IEEEbiography}
    [{\includegraphics[width=1in,clip,keepaspectratio]{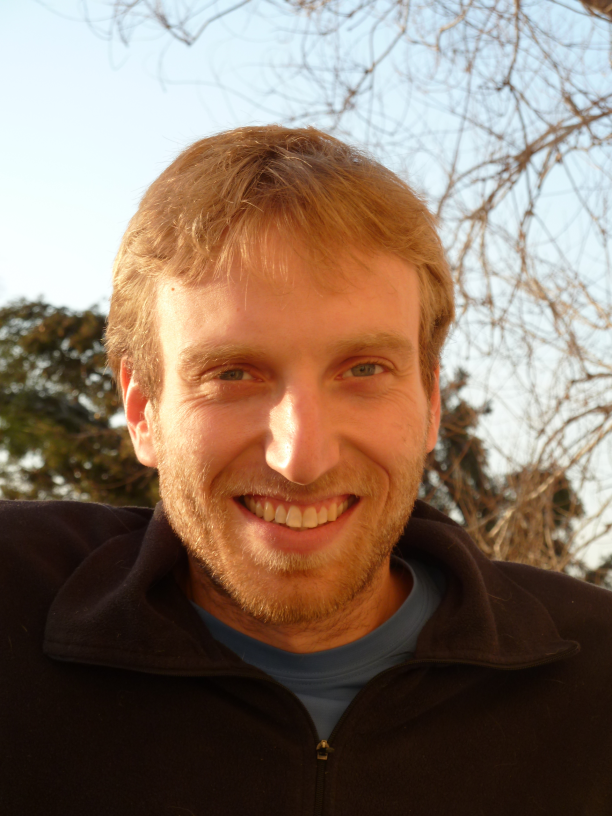}}]{Nir Shlezinger} (Member, IEEE) received his B.Sc., M.Sc., and Ph.D. degrees in 2011, 2013, and 2017, respectively, from Ben-Gurion University, Israel, all in electrical and computer engineering.
    From 2017 to 2019 he was a postdoctoral researcher in the Technion, and from 2019 to 2020 he was a postdoctoral researcher in Weizmann Institute of Science, where he was awarded the FGS prize for outstanding research achievements.
    He is currently an assistant professor in the School of Electrical and Computer Engineering in Ben-Gurion University, Israel.
    His research interests include communications, information theory, signal processing, and machine learning.
\end{IEEEbiography}

\begin{IEEEbiography}
    [{\includegraphics[width=1in,clip,keepaspectratio]{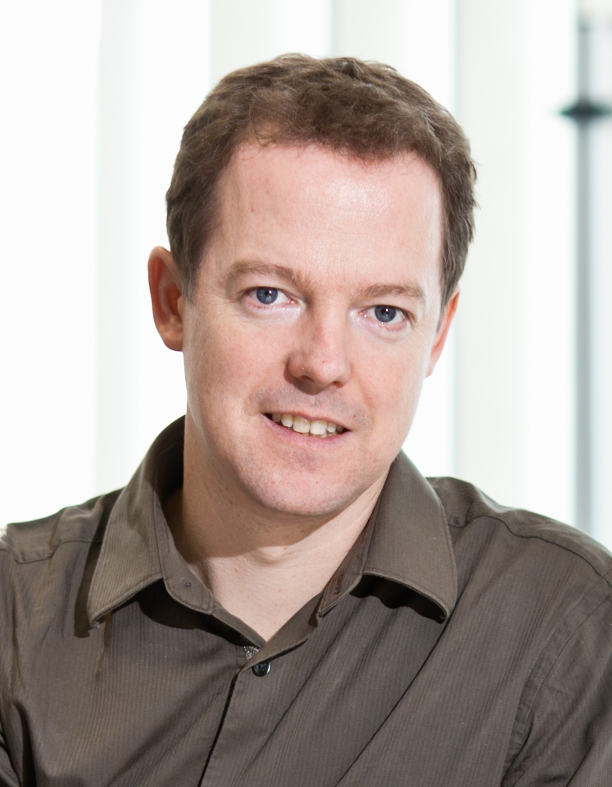}}]{Meik Dörpinghaus} (Member, IEEE) received the Dipl.-Ing. degree (with distinction) and the Dr.-Ing. degree (summa cum laude) both in electrical engineering and information technology from RWTH Aachen University, Aachen, Germany, in 2003 and 2010, respectively. From 2004 to 2010, he was with the Institute for Integrated Signal Processing Systems, RWTH Aachen University. From 2010 to 2013, he was a postdoctoral researcher at the Institute for Theoretical Information Technology, RWTH Aachen University. Since 2013, he has been a Research Group Leader at the Vodafone Chair Mobile Communications Systems and at the Center for Advancing Electronics Dresden (cfaed), Technische Universität Dresden, Germany. In 2007, he was a visiting researcher at ETH Zürich, Zürich, Switzerland. From 2015 to 2016, he was a visiting assistant professor at Stanford University, Stanford, CA, USA. His research interests are in the areas of communication and information theory.
    
    Dr. Dörpinghaus received the Friedrich-Wilhelm Preis of RWTH Aachen in 2004 and the Siemens Preis in 2004 for an excellent diploma thesis, and the Friedrich-Wilhelm Preis of RWTH Aachen in 2011 for an outstanding Ph.D. thesis. He has co-authored a paper that received a best student paper award at the IEEE Wireless Communications and Networking Conference 2018.
\end{IEEEbiography}

\begin{IEEEbiography}
    [{\includegraphics[width=1in,clip,keepaspectratio]{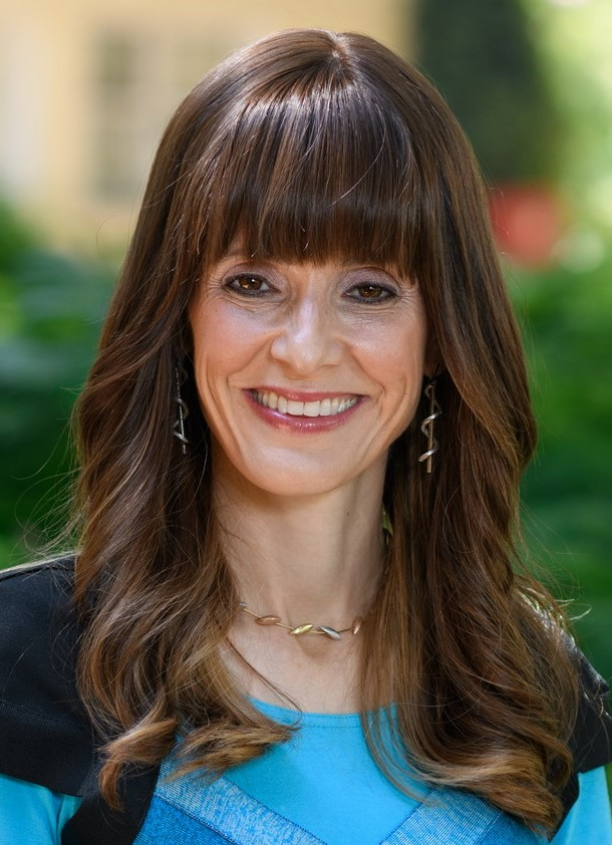}}]{Yonina C. Eldar} (Fellow, IEEE) received the B.Sc. degree in Physics in 1995 and the B.Sc. degree in Electrical Engineering in 1996 both from Tel-Aviv University (TAU), Tel-Aviv, Israel, and the Ph.D. degree in Electrical Engineering and Computer Science in 2002 from the Massachusetts Institute of Technology (MIT), Cambridge.

    She is currently a Professor in the Department of Mathematics and Computer Science, Weizmann Institute of Science, Rehovot, Israel. She was previously a Professor in the Department of Electrical Engineering at the Technion, where she held the Edwards Chair in Engineering. She is also a Visiting Professor at MIT, a Visiting Scientist at the Broad Institute, and an Adjunct Professor at Duke University and was a Visiting Professor at Stanford. She is a member of the Israel Academy of Sciences and Humanities (elected 2017), an IEEE Fellow and a EURASIP Fellow. Her research interests are in the broad areas of statistical signal processing, sampling theory and compressed sensing, learning and optimization methods, and their applications to biology, medical imaging and optics.

    Dr. Eldar has received many awards for excellence in research and teaching, including the  IEEE Signal Processing Society Technical Achievement Award (2013), the IEEE/AESS Fred Nathanson Memorial Radar Award (2014), and the IEEE Kiyo Tomiyasu Award (2016). She was a Horev Fellow of the Leaders in Science and Technology program at the Technion and an Alon Fellow. She received the Michael Bruno Memorial Award from the Rothschild Foundation, the Weizmann Prize for Exact Sciences, the Wolf Foundation Krill Prize for Excellence in Scientific Research, the Henry Taub Prize for Excellence in Research (twice), the Hershel Rich Innovation Award (three times), the Award for Women with Distinguished Contributions, the Andre and Bella Meyer Lectureship, the Career Development Chair at the Technion, the Muriel \& David Jacknow Award for Excellence in Teaching, and the Technion’s Award for Excellence in Teaching (two times).  She received several best paper awards and best demo awards together with her research students and colleagues including the SIAM outstanding Paper Prize, the UFFC Outstanding Paper Award, the Signal Processing Society Best Paper Award and the IET Circuits, Devices and Systems Premium Award, was selected as one of the 50 most influential women in Israel and in Asia, and is a highly cited researcher.

    She was a member of the Young Israel Academy of Science and Humanities and the Israel Committee for Higher Education. She is the Editor in Chief of Foundations and Trends in Signal Processing, a member of the IEEE Sensor Array and Multichannel Technical Committee and serves on several other IEEE committees. In the past, she was a Signal Processing Society Distinguished Lecturer, member of the IEEE Signal Processing Theory and Methods and Bio Imaging Signal Processing technical committees, and served as an associate editor for the IEEE Transactions On Signal Processing, the EURASIP Journal of Signal Processing, the SIAM Journal on Matrix Analysis and Applications, and the SIAM Journal on Imaging Sciences. She was Co-Chair and Technical Co-Chair of several international conferences and workshops. She is author of the book "Sampling Theory: Beyond Bandlimited Systems" and co-author of four other books published by Cambridge University Press.
\end{IEEEbiography}

\begin{IEEEbiography}
    [{\includegraphics[width=1in,clip,keepaspectratio]{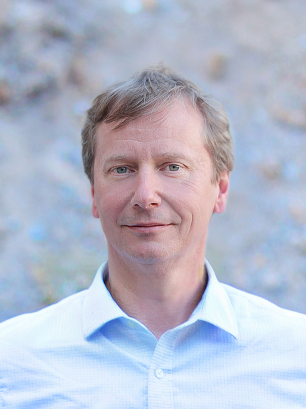}}]{Gerhard Fettweis} (Fellow, IEEE) is Vodafone Chair Professor at TU Dresden since 1994, and heads the Barkhausen Institute since 2018, respectively. In 2019, he was elected into the DFG Senate. He earned his Ph.D. under H. Meyr's supervision from RWTH Aachen in 1990. After one year at IBM Research in San Jose, CA, he moved to TCSI Inc., Berkeley, CA. He coordinates the 5G Lab Germany at TU Dresden. His research focusses on wireless transmission and chip design for wireless/IoT platforms, with 20 companies from Asia/Europe/US sponsoring his research.
    
    Gerhard is IEEE Fellow, member of the German Academy of Sciences (Leopoldina), the German Academy of Engineering (acatech), and received multiple IEEE recognitions as well as the VDE ring of honor. In Dresden his team has spun-out sixteen start-ups, and setup funded projects in volume of close to EUR 1/2 billion. He co-chairs the IEEE 5G Initiative, and has helped organizing IEEE conferences, most notably as TPC Chair of ICC 2009 and of TTM 2012, and as General Chair of VTC Spring 2013 and DATE 2014. 
\end{IEEEbiography}

\end{document}